\documentclass[a4, 11pt]{article}

\usepackage[authoryear,sort&compress]{natbib}
\usepackage{lineno,hyperref, algorithm, algorithmic}
\usepackage[utf8]{inputenc}
\usepackage{enumerate}
\usepackage{pgfgantt}
\usepackage{graphicx, booktabs}
\usepackage{placeins}
\usepackage{tikz}
\usepackage{geometry}
\usepackage{datetime}
\settimeformat{hhmmsstime}
\usepackage{amsmath, amssymb, bm, mathrsfs}
\usepackage{siunitx}
\usepackage{longtable,tabularx}
\usepackage{subcaption}
\usepackage{lipsum}
\usepackage{rotating}
\usepackage{authblk}

\usepackage{fancyhdr}

\modulolinenumbers[5]

\begin{document}
	\sloppy

	\title{Aircraft turnaround time estimation in early design phases: simulation tools development and application to the case of box-wing architecture \footnote{Accepted version of: M. Picchi Scardaoni, F. Magnacca, A. Massai, V. Cipolla, Aircraft turnaround time estimation in early design phases: Simulation tools development and application to the case of box-wing architecture, Journal of Air Transport Management, Volume 96, 2021, https://doi.org/10.1016/j.jairtraman.2021.102122.}}
	\author[1]{Marco Picchi Scardaoni }
	\author[2]{Fabio Magnacca}
	\author[1]{Andrea Massai}
	\author[1]{Vittorio Cipolla}
	\affil[1]{University of Pisa, Department of Civil and Industrial Engineering, Largo Lucio Lazzarino, 56122 Pisa, Italy}
	\affil[2]{University of Pisa, Department of Economics and Management, Via C. Ridolfi 10, 56124 Pisa, Italy}
		
	\maketitle

	\begin{abstract}
		This work deals with the problem of estimating the turnaround time in the early stages of aircraft design. The turnaround time has a significant impact in terms of marketability and value creation potential of an aircraft and, for this reason, it should be considered as an important driver of fuselage and cabin design decisions.   
		Estimating the turnaround time during the early stages of aircraft design is therefore an essential task. This task becomes even more decisive when designers explore unconventional aircraft architectures or, in general, are still evaluating the fuselage design and its internal layout.  In particular, it is of paramount importance to properly estimate the boarding and deboarding times, which contribute for up the 40\% to the overall turnaround time. 
		For this purpose, a tool, called SimBaD, has been developed and validated with publicly available data for existing aircraft of different classes. In order to demonstrate SimBaD capability of evaluating the influence of fuselage and cabin features on the turnaround time, its application to an unconventional box-wing aircraft architecture, known as PrandtlPlane, is presented as case study. Finally, considering standard scenarios provided by aircraft manufacturers, a comparison between the turnaround time of the PrandtlPlane and the turnaround time of a conventional competitor aircraft is presented. 
	\end{abstract}


	\maketitle
	
	\section{Introduction}
	The turnaround time (TAT) is the time period in which a landed aircraft is enabled to take-off for a new flight. According to \citep{IATA2017}, the standard turnaround process encompasses the following four major tasks: passenger and baggage handling, cargo and mail handling, load control and ramp handling. These tasks, which together can be referred to as \textit{aircraft ground handling}, define the servicing of an aircraft while it is on the ground, thus determining its TAT. This parameter is a key performance indicator of any aircraft. As a matter of fact, both airlines and airport management companies are interested in the TAT: it is one of the most important contractual items between the two parties, since it affects competitiveness and profitability \citep{More2014,Wu2000}.
	
	According to authors such as \cite{jenk1999}, the possibility to evaluate the TAT during the aircraft design phase, allows for the early assessment of its profitability. 
	Assuming an operating scenario where demand is unlimited and no competition exists, the best aircraft utilisation, i.e. the maximum number of flights an aircraft can perform per year, is determined by the TAT. The shorter is the TAT, the higher is the aircraft utilisation, the higher is the aircraft utilisation, the lower is the cost per available seat kilometre (CASK) that the airline bears to carry every single passenger on a certain route. This is because the amount of fixed direct operating costs (DOC) the airline bears to operate the aircraft is spread over an increased number of fares, hence kilometres and passengers \citep{mirza2008}.
	
	The efficiency of the aircraft turnaround is therefore an essential element of airlines success, especially for those regional and short-haul carriers \citep{belobaba2015} operating with a low cost business model \citep{lohmann2013} that requires no-frills services to meet the low fares target \citep{doganis2006} in a profitable way. In the perspective of the airline, the TAT represents hence a constraint that limits the value creation potential of the aircraft in its fleet.
	Moreover, \cite{Nyquist2008} found that the average cost to an airline company for each minute spent at the terminal is roughly $30\,\$$. Thus, each minute saved in the TAT of a flight has the potential to generate considerable annual savings. Reducing idle time on the ground will lead to improved airplane utilisation \citep{Qiang2014}. 
	
	When looking at the same problem through the lens of the airport, the TAT is again a constraint, since it limits the number of times per day the airport can sell the occupation of a given slot to airlines. The shorter is the TAT of the aircraft landing, the higher is the number of aircraft served. This not only translates into better aviation economic performance, due to the increased number of aircraft served in a certain time-frame, but also into better operational performance. Shorter TATs translate into higher gates, ground equipment and labour utilisation \citep{gillen2004} and, at the end, productivity. Nevertheless, higher productivity means also handling more passengers which in turn translates into higher non-aviation revenues potential. This is due to the increased number of people passing through the airport at a certain time. 
	
	However, estimating the TAT of a new aircraft design is not a straightforward task. The TAT results from a mechanism of interweaved activities with hierarchies. Among these activities, the ones depending on airport staff can be easily handled in the estimation effort, since airport staff is well trained to accomplish tasks following standard procedures in an efficient time. Whereas for such tasks standard duration used by aircraft manufacturers in Airport Planning manuals (e.g. Airbus and Boeing) can be considered, passengers boarding and deboarding operations need a different approach, since the large variance of their behaviour. A wide range of works on this aspect confirms the issues of simulating/evaluating boarding and deboarding phases \citep{Qiang2014, Gwynne2018} in the turnaround process and the consequent TAT estimation.
	
	The present paper aims to address the problem of estimating the TAT in the early stages of aircraft design, providing an approach which can be adopted in many cases, including non-conventional aircraft configurations.  For such cases, in fact, the conceptual and preliminary design phases are characterised by a higher level of uncertainty, which also touches the fuselage and cabin layout. The capability of predicting how the TAT is affected by the design choices can be used as a driver to move the design towards shorter TATs and with improved economic performances.
	
	In this context, an in-house simulation model, called SimBaD (Simulation of Boarding and Deboarding), has been developed for the estimation of boarding and deboarding times, considering new cabin layout aspects and deepening the modelling of those important phenomena such as seat interference.
	
	Although the TAT estimation approach here presented is valid for both conventional and non-conventional aircraft, its development has been motivated within the EU funded research project PARSIFAL (Prandtlplane ARchitecture for the Sustainable Improvement of Future AirpLanes, see \cite{PARSIFAL2017}), whose goal was the design and the economic impact assessment of a novel box-wing commercial aircraft architecture known as PrandtlPlane. 
	\begin{figure}[hbtp]
		\centering
		\includegraphics[width=2.5in]{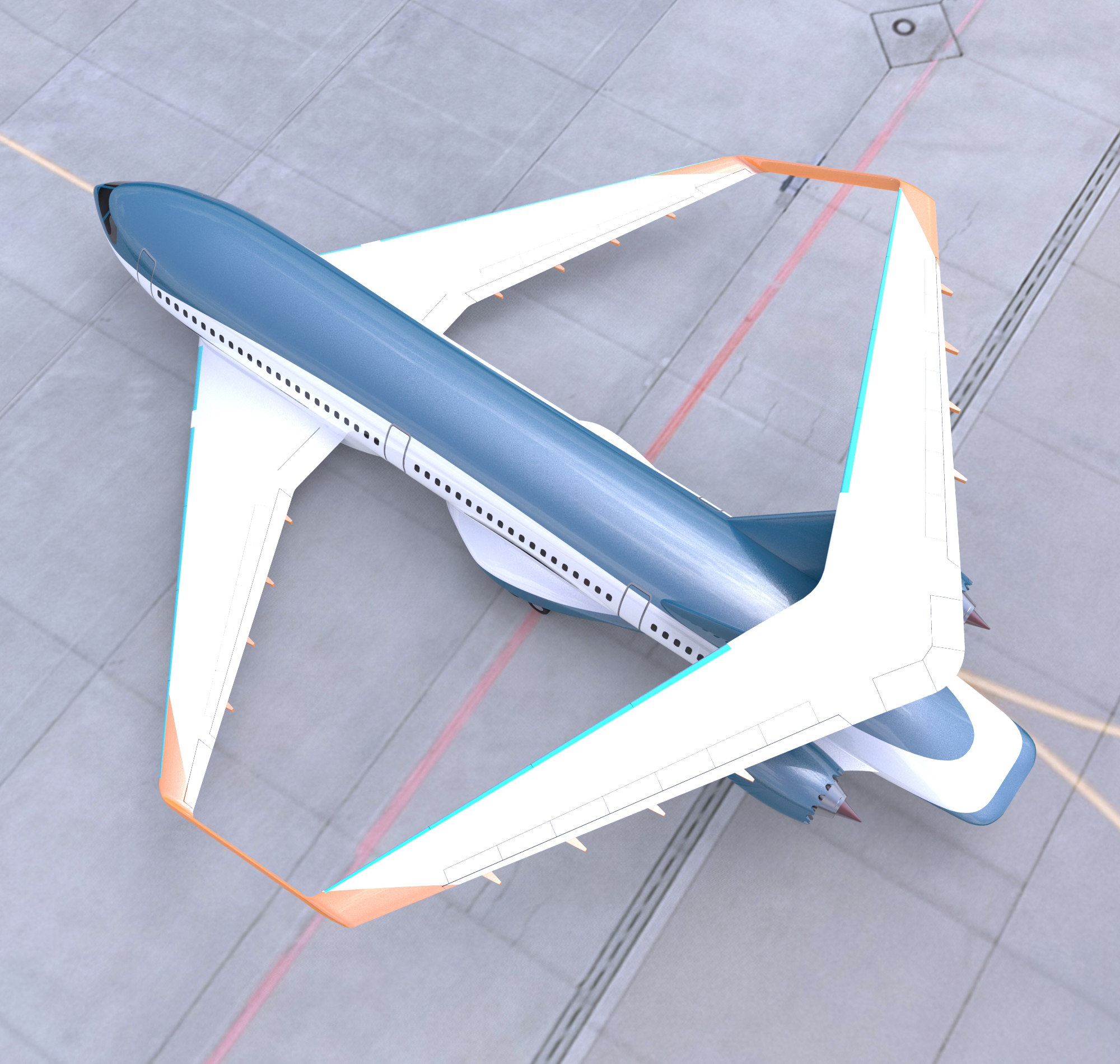}
		\caption{Artistic view of the PrandtlPlane object of study in PARSIFAL project \citep{PARSIFAL2017}}
		\label{fig:prp}
	\end{figure}
	As detailed in  \cite{Frediani2019, Cipolla2018b, Cipolla2018a}, this aircraft has been conceived to increase the payload capability up to about $310$ passengers, while keeping the same overall dimensions of today single aisle aircraft (e.g. Airbus 320 and Boeing 737 families), whose maximum number of seat is about $230$.
	In order to reach such a performance, the cabin has been conceived with a twin-aisle layout (see Fig. \ref{fig:2pax} in the following), with the peculiar characteristic of aisles wide enough to allow the simultaneous presence of two passengers (e.g. a passenger should be free to walk while another one is storing the luggage in the overhead bin). This feature has been introduced during the conceptual design phase in order to avoid the TAT increase associated to the larger number of passengers. 
	
	The goal of this work is twofold. Firstly, it aims at presenting a possible approach to estimate the TAT considering the effects of cabin layout on boarding and deboarding times. Secondarily, it aims at performing the assessment of the TAT for the PARSIFAL PrP case  in comparison to conventional competitors.
	
	The paper is structured as follows.
	Section \ref{sec:LR} presents the state of the art and the contribution of this work.
	Section \ref{sec:2} preliminary introduces the turnaround activities: some of them, emerge as a criticality for non-conventional aircraft, for which the available data may be null. 
	Then, Section \ref{sec:simbad} describes SimBaD as a tool for the simulation of boarding and deboarding phases. Section \ref{sec:tat} is dedicated to the estimation of the turnaround time of the case study of interest: the PARSIFAL PrP. Finally, Section \ref{sec:conclusions} contains some remarkable comments and prospects.

	\section{Literature Review and Scope of the Work}\label{sec:LR}
	In the literature, most of the studies focus on developing models for existing aircraft in order to propose new boarding strategies to reduce TAT. However, the attention is paid on boarding and deboarding phases. In any case, the literature seems more focused on comparative studies or on proposing more efficient boarding strategies for existing conventional aircraft, usually referenced as \textit{standard} aircraft (single-aisle with 3+3 seats abreast).
	A brief summary is presented in Tab.~\ref{tab:biblio}.
	
	In \cite{Landeghem2002}, the authors propose a numerical model to simulate different boarding patterns for a conventional aircraft, with the objective of understanding how to reduce the BT while increasing the quality perception of the passengers. The paper shows that any efficient boarding sequence should combine control over individual passengers. Based on this work, \cite{Ferrari2005} propose an agent-based boarding simulation for a conventional aircraft. As stated by the authors themselves, the absolute results from the simulations are meaningless: only comparisons are allowed. Moreover, the seat interference is evaluated through a prolonged occupancy of the aisle.\\
	In \cite{Bazargan2007}, it is proposed a mixed-integer programming model to reduce the number of passenger interferences in a conventional cabin during boarding phases. It is claimed that analytical approaches can reveal more efficient boarding strategies. Several congestions events are taken into account through simplified analytical relationships.\\
	In \cite{Qiang2014}, a boarding model based on agent-based cellular automata, widely used in traffic flow theory, is proposed to demonstrate the efficiency of new boarding strategies based on passengers' individual properties, for a standard aircraft.\\
	In \cite{Milne2014}, the authors investigate the impact of passenger luggage on boarding time only for a conventional cabin layout using an agent-based framework. The objective of this work is to assign passengers to seats, so that luggage is distributed evenly throughout the plane, thus achieving time saving.
	Similarly, \cite{Milne2016, Milne2018} propose a two-step mixed-integer programming approach to reduce BT for a conventional aircraft. Based on the amount of passengers' pieces of luggage, the first step assigns seats to passengers, so to minimise the BT. The second step assigns again seats to passengers while constraining the BT to be less than that determined in the first step.\\
	In \cite{Qiang2016}, a numerical asymmetric simple exclusion process is applied to the boarding process. They distinguish three regimes of boarding, depending on the  passenger arrival rate and luggage disposal rate. Even though the model catches the transitions between these regimes, the model provides nearly no analytic discussions. As stated by the authors themselves, it is a major drawback, since it cannot lead to a deeper understanding of the boarding problem.
	In \cite{Qiang2016a}, an agent-based model is used to propose a strategy with group behaviour to study boarding and deboarding. In fact, the authors claim that potential optimisation might be achieved by considering the boarding and deboarding processes in a integrated way. Some structured deboarding strategies, related to boarding strategies, are also suggested in this work.  \\
	In \cite{Zeineddine2017}, a new boarding strategy is proposed. It makes efficient use of the available technology (electronic notification systems for disseminating boarding information to passengers through their personal devices) to support the boarding process, allows passengers to board in cliques to their seats. Simulations through an agent-based approach focus on conventional cabin layouts.\\
	The attention to new features of novel aircraft configurations, in the turnaround time perspective, is addressed in \cite{Schmidt2016}, with a focus on electric and hybrid aircraft. The authors develop an analytical model to quantify the effect of battery change or recharge in the TAT of such aircraft. Moreover, they wonder if cabin layouts could enable faster passenger egress and ingress.\\
	An agent-based framework is also proposed in  \cite{Jafer2017} to compare eight different boarding strategies, so to chose the best one. The proposed model suffers of the lack of an explicit simulation of seat interference, and the results seem to present over-conservative estimation of boarding time.\\
	\cite{Tang2018} investigate the groups behaviours and the quantity of luggage for efficient boarding phases. The authors assess a differential analytical model.	However, such a model presents many limits, as recognised by the authors themselves. For instance, it is not based on experimental/empirical data and each passenger’s preference for seat is not considered.\\
	In \cite{Schmidt2017}, authors focus on the cabin design of a small regional aircraft, with the aim of reducing the impact of boarding and deboarding time within the turnaround process. Simulations are performed through an agent-based tool. The investigation is mainly focused on door positions and seating concepts.
	With a similar approach, \cite{Schmidt2017a} aim at identifying reduction potentials for conventional single- and twin- aisle aircraft, acting on the design of the internal cabin. The results are used within an operational assessment of the aircraft. The simulations are focused on the effect of foldable seats. Results show that cabins which can be adapted during boarding could provide flexible cabin design which can be modified depending upon flight phase requirements.\\
	The impact of new seat design and boarding methodologies, such as the Side-Slip Seat, Flying Carpet, Foldable Passenger Seats on the boarding phases is similarly and deeply investigated in \cite{Schultz2017a, Schultz2018b}. The results are aligned to the previous ones.
	In \cite{Schultz2019} an approach for a real-time evaluation of the aircraft boarding progress is presented. The aim is to show an approach for a real-time evaluation of the aircraft boarding progress. A simulation environment is also used to show both the common progress of aircraft boarding and the capability of predicting the boarding time under different levels of uncertainty.
	This information is used as an input to derive new real-time evaluation metrics. Such indicators are then implemented in a hardware prototype environment and used in field trials with promising predictive results. Such metrics have also been used to train an Artificial Neural Network for the prediction of boarding times of a conventional cabin layout. The training of the Neural Network has been performed via stochastic, forward-directed, one-dimensional and discrete process. In the numerical simulations, the passenger movement only depends on the state of the next cell, and pre-established instructions are prescribed to passengers.\\
	{ More recently, due to the Covid-19 pandemic, some works such as \cite{Milne2021, Islam2021, Cotfas2020, Salari2021} have focused on correlations between boarding strategies and virus spreading, suggesting new boarding strategies, new sitting policies and social distancing to contrast the disease diffusion.}\\

	It seems evident that the current literature background is affected by the following limitations: (i) the comparative approaches allows for less accurate models, since only differences between measured times (and not punctual estimations) are needed; (ii) models and simulations regard mainly \textit{standard} narrow-body aircraft, whilst the attention to novel aircraft configuration is rarely addressed; 
	(iii) the focus is mainly on the boarding phase; (iv) seat interference is barely explicitly modelled; (v) the cabin layout is considered most of the time as given: simulations are rarely exploited as a tool to design the cabin interior; (vi) the turnaround time is rarely considered in the preliminary design phases.\\	
	{ Moreover, apart from purely numerical and Machine Learning approaches, from the above literature survey the most common approach seems to be the use of agent-based models to predict the boarding/deboarding times. Agent-based modelling simulates the actions and interactions of independent agents in order to figure out the behaviour of a system, but the common feature of such works is that the behaviour of the agents (the passengers) is often simplified in some fundamental aspects, such as seat interference, and does not take  the possibility of overtaking in the aisles into account.}


	The present paper stands to contribute to the existing state of the art by addressing the problem of TAT estimation in the preliminary design phases of an aircraft, conventional or not. 
	Accordingly, this work aims at providing accurate and punctual estimations of TAT, rather than values that are meaningful only in relative terms. Naturally, comparative studies are still possible.
	To estimate both boarding and deboarding times, SimBaD has been developed, which allows to evaluate different internal cabin layouts. 	
	It is noteworthy that SimBaD explicitly models seat interference congestions and other meaningful sources of delay, as will be detailed.
	Finally, to demonstrate the applicability of the proposed approach, a case study has been conducted on the TAT estimation of a novel aircraft architecture, the PrandtlPlane.

	\begin{sidewaystable}
		\small
		\centering
		\begin{tabularx}{\textwidth}{XllllXX}
			\toprule
			Reference & BT & TAT & Layout & N° doors & Goal & Methodology\\
			\midrule
			\cite{Landeghem2002, Ferrari2005} & BT & No & Conv. &1 &Understanding how to reduce the total boarding time while increasing the quality perception of the passengers. & Numerical model and cell-based simulations.\\
			\midrule
			\cite{Bazargan2007} & BT & No & Conv. & 1 & Generate new efficient boarding strategies. & Mixed-integer linear programming.\\
			\midrule
			\cite{Qiang2014} & BT& No& Conv. &1 & Provide a more efficient boarding strategy in consideration of passengers' individual properties. & Cellular automaton model. \\
			\midrule
			\cite{Milne2014} & BT& No&Conv. &1 &Investigate the impact of passenger luggage on BT &Discrete Event Simulations\\
			\midrule
			\cite{Milne2016, Milne2018} & BT& No& Conv.&1 &Provide a more efficient boarding strategy in consideration of passengers' pieces of luggage. & Mixed-integer programming.\\
			\midrule
			\cite{Qiang2016} & BT&No &Conv. &1 &Investigate the relationship between passenger arrival time interval and airplane BT. &Numerical asymmetric simple exclusion process.\\
			\midrule
			\cite{Qiang2016a} &BT, DT & No& Conv.&1 &Study the boarding and deboarding processes in an integrated way. Propose structured deboarding strategies.  &Cellular automaton model. \\
			\midrule
			\cite{Zeineddine2017} & BT &No &Conv. &1 &Proposing a new boarding strategy making efficient use of the available technology to support the boarding process  & Discrete Event Simulations\\
			\midrule
			\cite{Schmidt2016, Schmidt2017, Schmidt2017a} & BT, DT& Yes & {(Un)Conv.} &1 & Investigate the effect of hybrid or electric propulsion on TAT. Investigate door position and foldable seats effects on regional and mid-range aircraft. & Agent-based tool.\\
			\midrule
			\cite{Jafer2017} & BT & No & Conv. & 1  & Compare eight boarding strategies. & Cellular Discrete-Event System Specification modelling.\\
			\midrule
			\cite{Tang2018} & BT & No & Conv. & 1 & Quantify the effect of group behaviour and of quantity of luggage.&  Numerical differential model.\\
			\midrule		
			\cite{Schultz2019, Schultz2018b, Schultz2017a} & BT & No & Conv. & 1, 2 & Predict the progress of a running boarding event. New indicators for enabling a real-time evaluation of the aircraft boarding progress. Assessment of the effects of Flying Carpet, Side-Slip Seat and Foldable Passenger Seat.& Agent-based tool, Artificial Neural Networks\\
			\bottomrule
		\end{tabularx}
		\caption{Comparative summary of the literature}
		\label{tab:biblio}
	\end{sidewaystable}

	\section{The Turnaround Process}\label{sec:2}
	The standard turnaround process encompasses different airport activities that involve the logistics of the airport as a whole: passenger and baggage handling, cargo and mail handling, load control, ramp handling \citep{IATA2017}. These activities follow a chronological strict time line and some of them are executed concurrently. For every specific aircraft type, aircraft manufacturer manuals provide a precise outline of such chronological sequence of tasks, displaying the activities of interest together with their \textit{typical time} during aircraft turnaround. These activities include the following: 
	\begin{enumerate}[(a)]
		\item \textbf{Passengers handling}, which includes deboarding, boarding, headcounting from the last passenger seating (LPS) time;
		\item \textbf{Cargo handling}, which includes baggage and freight unloading and loading in the cargo bays, generally divided in aft and fore bays;
		\item \textbf{Refuelling};
		\item \textbf{Cleaning};
		\item \textbf{Catering}, which includes supply of provisions, potable water, etc.
	\end{enumerate}
	Each activity is anticipated by the equipment positioning (e.g., stairs, cargo loaders, belt loaders, etc.) that, evidently, is necessary to carry out the considered activity, and it is followed by the equipment removal once the activity is concluded.
	The total time allocated to complete an activity is defined \textit{engagement time} \citep{IATA2017}. Within the turnaround process, however, some activities are considered as critical. They must be concluded before the following activities can start. The combination of such critical (i.e. dependent) activities, whose succession takes the greatest time to complete, determines the so-called \textit{critical path}. As a result, given a certain business model of the carrier, the TAT depends on the duration of the critical path, and modifications of the engagement time of the critical activities directly affect the aircraft minimum TAT.
	
	During the design phase of a new aircraft, an important step is the time estimation of the aforementioned critical activities. Some of them are not influenced by the specific aircraft architecture, at least for classes of aircraft. For these activities, data from existing aircraft manuals can be assumed in the preliminary design phases. As an example, one can assume that containers are loaded and unloaded in a time duration that is independent from the aircraft configuration (but may depend on the container type). Similarly, refuelling depends only on the fuel flow and not on the aircraft architecture. Moreover, catering time is proportional to the number of passengers. As for equipment positioning and removal, airport facilities are standardised worldwide. They accommodate different aircraft that, in turn, are designed to be compliant with airport facilities worldwide. Once activities durations are known for a particular aircraft of the same class, the estimation of the engagement time of the activity considered for the aircraft at hand is easily done.
	
	The highest source of uncertainty when estimating the TAT is embedded in the activity of passenger handling, because of the high standardisation of the air transportation industry and, more precisely, of ground handling activities. As a matter of fact, different cabin layouts do affect the modality by which passengers embark and disembark. Since boarding and deboarding are likely to lie in the critical path, an estimation error propagates directly to the overall TAT. It is than crucial to have a tool to estimate such likely-critical activities for the TAT, since the preliminary design phases.
	

	\section{SimBaD Tool to Simulate Boarding and Deboarding Phases}\label{sec:simbad}
	SimBaD (Simulation of Boarding and deboarding) is an in-house tool developed in order to explicitly simulate the boarding and deboarding phases for unconventional cabin layouts.
	It is coded in Python language, using only standard libraries. The approach is to model passengers as Finite-State Machines (FSMs), who can move and act in a discrete internal cabin layout, subdivided into elementary cells. The next-step \textit{State} of a cell depends on both its actual State and the States of the neighbour cells.  Each simulation step corresponds to a clock time, chosen as one-tenth of a second.
	
\begin{figure}[hbtp]
		\centering
		\subfloat[B737 cabin discretisation]{\includegraphics[width=0.48\columnwidth]{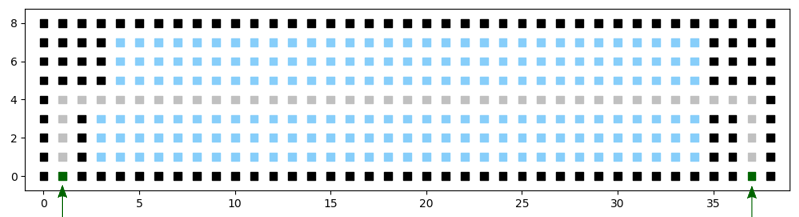}\label{fig:b737disc}}%
		\subfloat[A330 cabin discretisation]{\includegraphics[width=0.48\columnwidth]{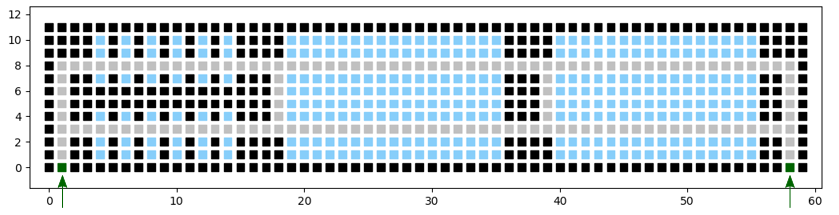}\label{fig:a3302Ddisc}}%
		\hfill
		\subfloat[PARSIFAL PrP cabin discretisation]{\includegraphics[width=0.48\columnwidth]{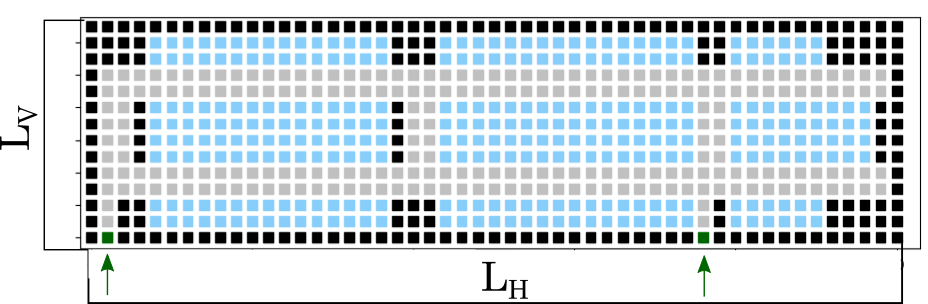}\label{fig:prpdiscrete}}
		\caption{Examples of cabin discretisation (active doors are highlighted with arrows)}
		\label{fig:discret}
	\end{figure}

	As an example, some cabin discretisations are shown in Fig.~\ref{fig:discret}.
	Four are the main cell types, highlighted in different colors: the walls (black), which represent non-crossable entities, the empty seats (light-blue), the aisles (grey), and the active doors (green). The cabin discretisation is a basilar input for SimBaD; the geometry can be easily defined in an \textit{ad-hoc} formatted text file. The user is free to provide a generic cabin layout, which can be easily modified.
	
	Passengers are modelled as a FSMs using Python Classes.
	Formally, each passenger is an entity of the type
	\begin{equation}\label{eq:pax}
		\mathscr{P}\left(n_{\mathrm{seat}}, t_{\mathrm{lug}}, v_\mathrm{H}, t_{\mathrm{door}}, \mathrm{ID}_{\mathrm{door}}, \mathrm{IF}\right),
	\end{equation}
	where $n_{\mathrm{seat}}$ is the passenger' seat number, $t_{\mathrm{lug}}$ is the time for storing (recuperating) the luggage, $v_\mathrm{H}$ is the velocity of the passenger along corridors,  $\mathrm{ID}_{\mathrm{door}}$ is the identification number of the entrance (exit) door, $\mathrm{IF}\,\in\,[0, 1]$ is the interference factor, associated to the probability of overtaking (further comments will follow).
	If $n_{\mathrm{pax}}^{\mathrm{max}}$ is the maximum number of passengers of the aircraft at hand, SimBaD simulations can involve a fraction of them, i.e. $n_{\mathrm{pax}} = \mathrm{LF}\, n_{\mathrm{pax}}^{\mathrm{max}}$, where $\mathrm{LF}\,\in\,[0, 1]$ is the load factor, which measures the crowdedness of the cabin.
	
	Algorithm \ref{alg:general} shows the conceptual macro structure of SimBaD.
	\begin{algorithm}
		\caption{SimBaD scheme}
		\label{alg:general}
		\begin{algorithmic}
			\REQUIRE Cabin discretisation file
			\REQUIRE $0\leq \mathrm{IF} \leq1$
			\REQUIRE $0\leq \mathrm{LF} \leq1$
			\FOR{$i=1, \cdots, n_{\mathrm{pax}}$}
			\STATE initialise $\mathscr{P}_i$ 
			\ENDFOR
			\STATE $t \leftarrow 0$ 
			\WHILE{\NOT all $\mathscr{P}_i$ seated (debarked)}
			\STATE $t \leftarrow t+1$
			\FOR{$i=1$ \TO $n_{\mathrm{pax}}$}
			\STATE update $\mathscr{P}_i$
			\ENDFOR
			\ENDWHILE
			\RETURN $t$
		\end{algorithmic} 
	\end{algorithm}
	Inputs are the cabin discretisation file, the interference factor $\mathrm{IF}$ and the load factor $\mathrm{LF}$. Successively,  passenger Objects are initialised: to each of them are assigned $\{n_{\mathrm{seat}}, t_{\mathrm{lug}}, v_\mathrm{H}, t_{\mathrm{door}}, \mathrm{ID}_{\mathrm{door}}\}$, as discussed for Eq. \eqref{eq:pax}. Hence, clock time is set to zero and passengers are all updated in their States, at each time clock. Once the exit condition is satisfied, i.e. all passengers are sat in the boarding case or all passengers are debarked in the deboarding case, the simulation arrests, and the final resulting time is returned as output.
	It is noteworthy that SimBaD can deal with different boarding strategies: Random, Outside-In, Back-To-Front, Rotating-Zone, User-defined. As for parameters $t_{\mathrm{lug}}$, $v_\mathrm{H}$, $t_{\mathrm{door}}$, they can be kept constant for every passenger or they can be assigned according to a certain statistical distribution.
	In particular, throughout the paper, in the spirit of \citep{Schultz2018}, Weibull's probability density functions (PDFs) have been considered, of the form
	\begin{equation}
		f(x; \alpha, \beta, \theta) = \begin{cases}
			\dfrac{\alpha}{\beta}\left(\dfrac{x-\theta}{\beta}\right)^{\alpha-1}\textrm{e}^{-\left(\frac{x-\theta}{\beta}\right)^\alpha}, & x\geq \theta,\\
			0, & x< \theta, 
		\end{cases}
	\end{equation}
	where $\alpha$ is the shape parameter, $\beta$ the scale parameter and $\theta$ is the offset parameter of the PDF.
	
	\subsection{SimBaD Parameters}
	As stated before, the aircraft internal layout is considered as discrete: to each unitary cell is associated a State; the principal ones are collected in Tab.~\ref{tab:states}.
	\begin{table}
		\centering
		\caption{Principal SimBaD cells States}
		\label{tab:states}
		\begin{tabular*}{\textwidth}{ll}
			\toprule
			{State} &  {Value}\\
			\midrule
			{Empty seat} & {1}\\
			{Empty aisle} & {0}\\
			{Wall} &  {-1}\\
			{Occupied aisle} &  {-2}\\
			{Occupied seat} &  {-3}\\
			{Active Door}&  {-5}\\
			\bottomrule
		\end{tabular*}
	\end{table}
	These parameters make possible the identification of the state (or nature) of the cells, for every time step, from the algorithm viewpoint. The initial layout of the cabin, i.e. the empty cabin, must be provided by the user. As already said, it can be done by specifying in a proper formatted file the coordinates of the wall-cells, of the seat-cells, of the aisle-cells and of the door-cells. During the simulation, the states of the cells are updated. For instance, if an empty aisle-cell is occupied by a passenger, the state will be updated from $0$ to $-2$. 
	
	\color{black}
	With reference to Fig.~\ref{fig:prpdiscrete}, let $L_H$ and $L_V$ be the cabin planform overall dimensions, and let $n_H$ and $n_V$ be the number of cells in the respective directions. Therefore, $u_H:=L_H/n_H$ and $u_V := L_V/n_V$ are the unit length in the horizontal and vertical directions, respectively. In general, $u_H \neq u_V$, which means that the discretisation scales differently the two directions.
	To eliminate this inconsistency, introducing the parameter $\gamma := u_H/u_V$, which depends only on the overall geometry of the cabin, one can modify the velocity of passengers moving in the vertical direction.
	Looking at Alg.~\ref{alg:general}, it is more convenient to express the velocity of a passenger as the time an aisle cell is occupied, in order to have a direct link with the clock time of the simulation.
	Therefore, let $t_H:=u_H/v_H$ and $t_V := u_V/v_V$ be the times of occupancy of a cell associated to a passenger moving in the vertical or horizontal direction, respectively. Posing $u_V := u_H/\gamma$, since the aim of restoring the physical condition $v_H = v_V$, one obtains $t_V = t_H/\gamma$. In the following, $u_H$ is assumed as the reference cell unit length.
	
\begin{figure}[hbtp]
		\centering
		\subfloat[$t_{lug}$]{\includegraphics[width=0.45\textwidth]{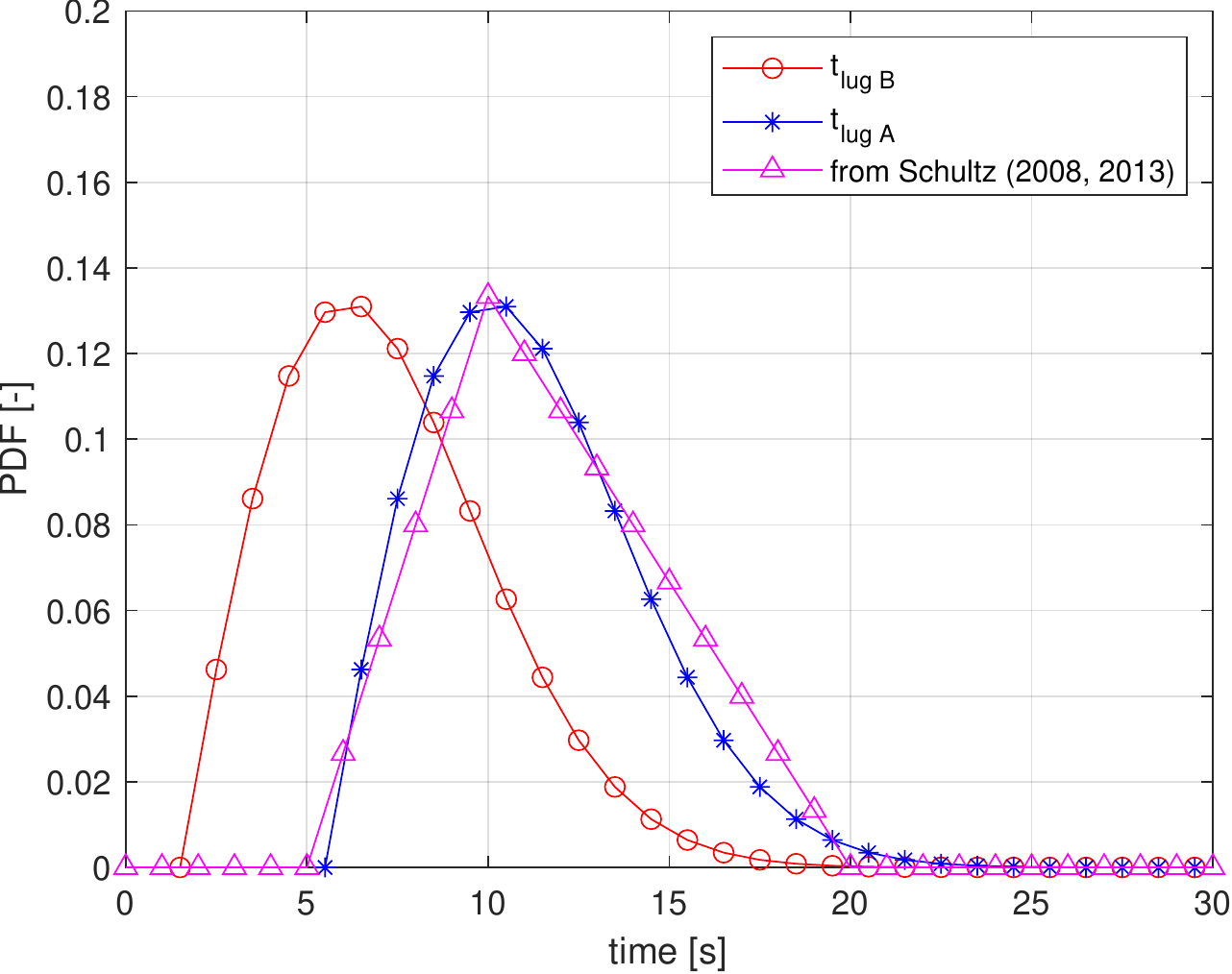}\label{fig:weib}}\hfill
		\subfloat[$t_{H}$]{\includegraphics[width=0.45\textwidth]{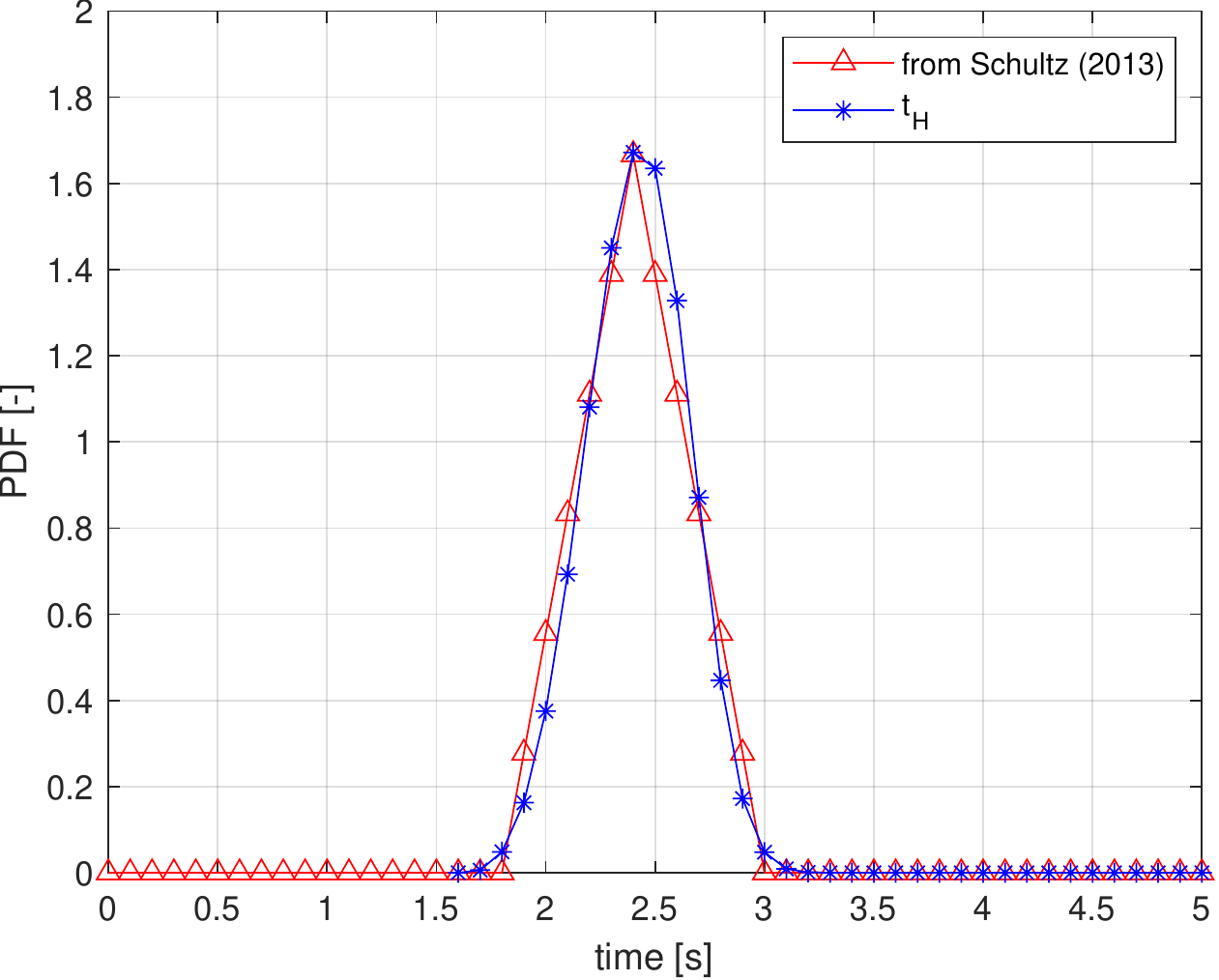}\label{fig:weibh}}\hfill
		\caption{Weibull distributions}
		\label{fig:weibullsssss}
	\end{figure}

	Table \ref{tab:weib} reports the Weibull parameters used for the distributions of $t_H$ and of $t_{\mathrm{lug}}$ associated to two luggage politics. 
	As far as luggage politic is concerned, condition \textit{A} considers every passenger to have a piece of luggage to store, e.g. a cabin trolley. Conversely, condition \textit{B} considers the possibility to have a consistent amount of passengers having small luggage or small bags that can be put below the seat, resulting in a short time of aisle occupancy. The latter luggage politic is associated to economic fares. In any case, the cell dedicated to the luggage storage/recovering is the one just in front of the passenger seat row.\\
	The PDFs for politics \textit{A} and \textit{B} are plotted in Fig. \ref{fig:weib}. 
	Luggage politics of type \textit{A} corresponds to the piecewise-linear distribution proposed by \cite{Schultz2013, Landeghem2002}, where each passenger is considered to store a piece of luggage. The counterpart for politics \textit{B} is obtained by a translation of the distribution \textit{A}. In so doing, the shape of the distribution is preserved, but the minimum time to store the luggage shifts from $5.5\,\si{\second}$ to $1.5\,\si{\second}$. Note that this value is conservative: in low cost fares there are passengers with small luggage which do not occupy the aisle at all (the corresponding occupancy time would be $0\,\si{\second}$). \\
	The distribution for $t_{H}$ is associated to the piece-wise distribution from \cite{Schultz2013, Landeghem2002}, as depicted in Fig. \ref{fig:weibh}.
	\color{black}
	\begin{table}
		\centering
		\caption{Weibull parameters}
		\label{tab:weib}
		\begin{tabular*}{\textwidth}{llll}
			\toprule
			{Variable} &  {$\alpha$} & {$\beta$} & {$\theta$}\\
			{} &  {$[-]$} & {$[\si{\second}]$} & {$[\si{\second}]$}\\
			\midrule
			{$t_H$} & {$0.9$}& {$4$}& {$1.6$}\\
			{$t_{lug\;A}$} & {$2$}& {$6.5$}& {$5.5$}\\
			{$t_{lug\;B}$} & {$2$}& {$6.5$}& {$1.5$}\\
			\bottomrule
		\end{tabular*}
	\end{table} 
	The $t_{\mathrm{door}}$ parameter, which simulates the time spent at the entrance door for showing the ticket to the airline personal, is considered constant, equal to $2\,\si{\second}$. Naturally, this parameters is effective only in the boarding phase.
	
	\subsection{Path Self-determination}
	As a general consideration, not-imposed passengers trajectories, which may evolve by themselves during the simulation, are a desideratum requirement.
	Moreover, some new concept of aircraft proposes wider aisles, which introduce the problem of simulating 2D trajectories. In fact, passengers are allowed to move in the width direction of aisles and, if necessary, to overtake slower or stationary passengers. 
	
\begin{figure}[hbtp]
		\centering
		\includegraphics[width=2in]{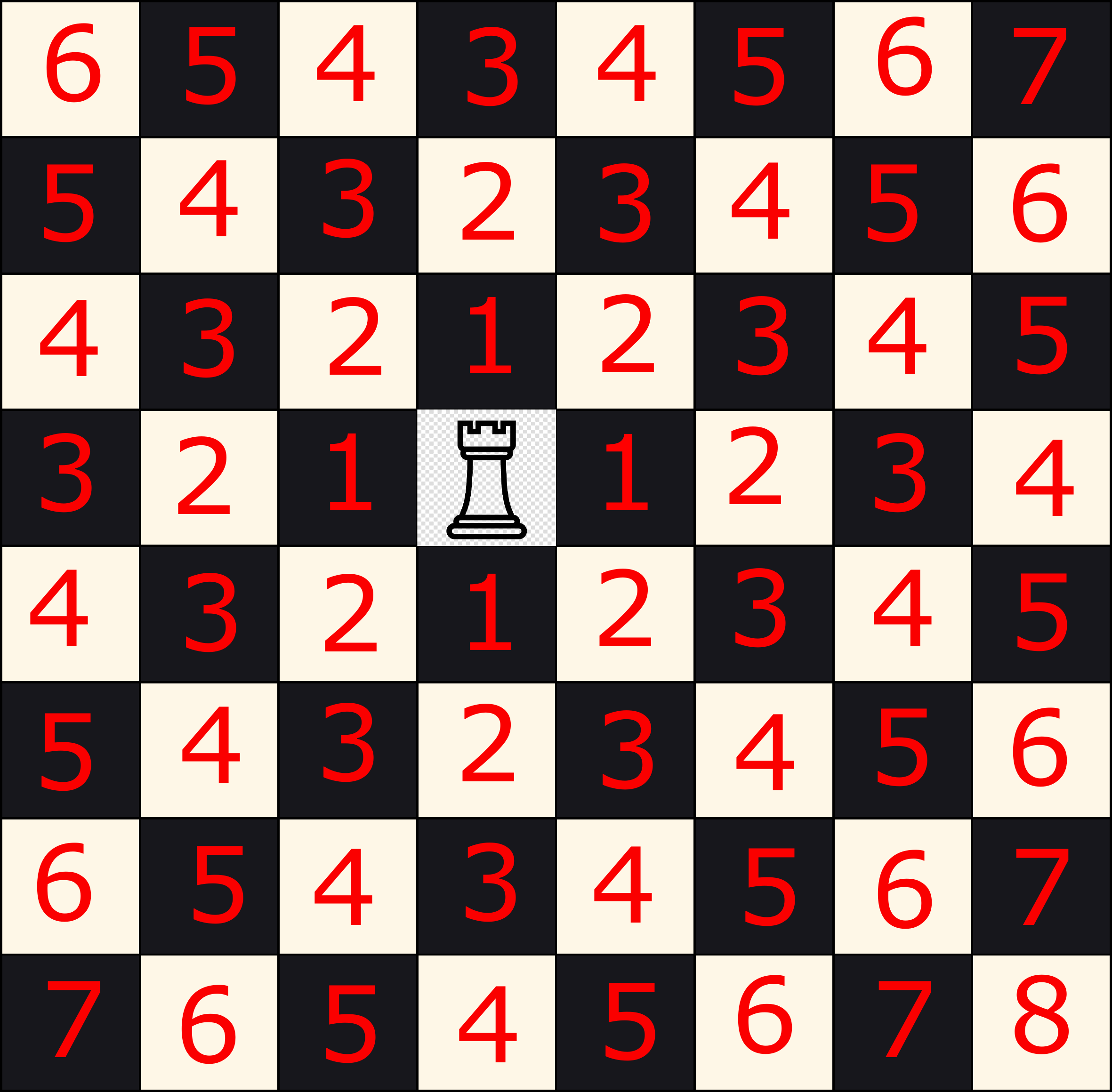}
		\caption{Locus of cells having the Manhattan distance equal to the red number within the cells, measured from the cell occupied by the Rook.}
		\label{fig:chessboard}
	\end{figure}

	SimBaD implements an algorithm based on a particular version of the Chebychev norm, called Manhattan distance (or Taxicab metric).
	Formally, it is defined as the $L^1$ norm $\mathscr{D}$ between two generic vectors $\textbf{p}$ and $\textbf{q}$. In a two-dimensional space, it reads
	\begin{equation}\label{eq:manhattan}
		\mathscr{D}\left(\textbf{p},\textbf{q}\right) = \sum\limits_{i=1}^2\left|p_i - q_i \right|.
	\end{equation}
	More intuitively, the Taxicab metric equals the number of moves that the Rook, in the Chess game, needs to do to pass from the actual cell to another one, as shown in Fig.\ref{fig:chessboard}.
	In SimBaD, passengers are allowed to move only in the four cardinal directions (N, E, S, W), associated to steps in direction $(0,1)$, $(1,0)$, $(0,-1)$, $(-1,0)$ respectively, or to stay in their actual cell (0), associated to the null step $(0,0)$. Furthermore, passengers can move only in the direction of their target seat. This assumption avoids counter-moving passengers with respect to the main passenger flow direction.
	\begin{algorithm}
		\caption{Next-step algorithm (basic)}
		\label{alg:nextstep}
		\begin{algorithmic}
			\REQUIRE Actual position $\textbf{p}$
			\REQUIRE Target position $\textbf{q}$
			\REQUIRE Cabin grid States
			\STATE $\textbf{v}\leftarrow \{\}$
			\STATE $\textbf{c} = \{(0,1), (1,0), (0,-1), (-1,0)\}$
			\FOR{$i=1$ \TO $4$}
			\STATE $\textbf{r} \leftarrow \textbf{c}[i]$ 
			\IF{passenger in $\textbf{p}$ can move and cell $\textbf{p}+\textbf{r}$ is empty}
			\STATE $d \leftarrow \mathscr{D} (\textbf{p} + \textbf{r},\textbf{q})$
			\STATE $\textbf{v}\leftarrow \textbf{v}.\mathrm{append}(\{d, \textbf{r}\})$
			\ENDIF
			\ENDFOR
			\IF{length($\textbf{v}$)$ > 0$}
			\STATE $\textbf{v}\leftarrow$ $\textbf{v}.\mathrm{sort}(d)$
			\RETURN next step $\leftarrow\textbf{v}[1,2]$
			\ELSE
			\RETURN next step $\leftarrow \textbf{p}$
			\ENDIF
		\end{algorithmic} 
	\end{algorithm}
	Algorithm \ref{alg:nextstep} describes the core of the determination of the next-step direction. For a passenger who can move, Alg. \ref{alg:nextstep} evaluates the Manhattan distances (Eq. (\ref{eq:manhattan})) between the target seat and the four cells surrounding the current one, in direction of the four cardinal points. Cells that can accept the incoming passenger are stored, together with the associated Manhattan distance. 
	If there exists at least one cell that can accept the passenger, the algorithm returns the cell which minimises distance from the target seat cell. In the next step, the passenger will occupy the cell. Conversely, if no cell exists, the passenger remains in the current cell also in the next clock time.
	
	With this assumptions, a passenger in an aisle having at least two cells in its transversal direction, having an \textit{obstacle} in front of him, will move in another cell of the aisle preserving the column, i.e. in direction N or S\footnote{Note that moving backward is not allowed, to avoid counter-moving passengers.}. In the following clock times, the passenger will overtake the obstacle.
	
	The possibility of overtaking can be allowed with restrictions. 
	As stated before, the interference factor parameter $\mathrm{IF}$ has been introduced  to control the probability of overtaking. Assuming a multi-cell width aisle, $\mathrm{IF}=0$ means that a passenger can always overtake another passenger. Conversely, $\mathrm{IF}=1$ means that no overtaking can happen. The former situation takes advantage of the wider aisle, whilst in the latter the advantage to have a wider aisle is null. A more likely situation is described by a value of the $\mathrm{IF}$ between these two limit cases.

	\subsection{Seat Interference}
\begin{figure}[hbtp]
		\centering
		\subfloat[Type 1]{\includegraphics[height=1in]{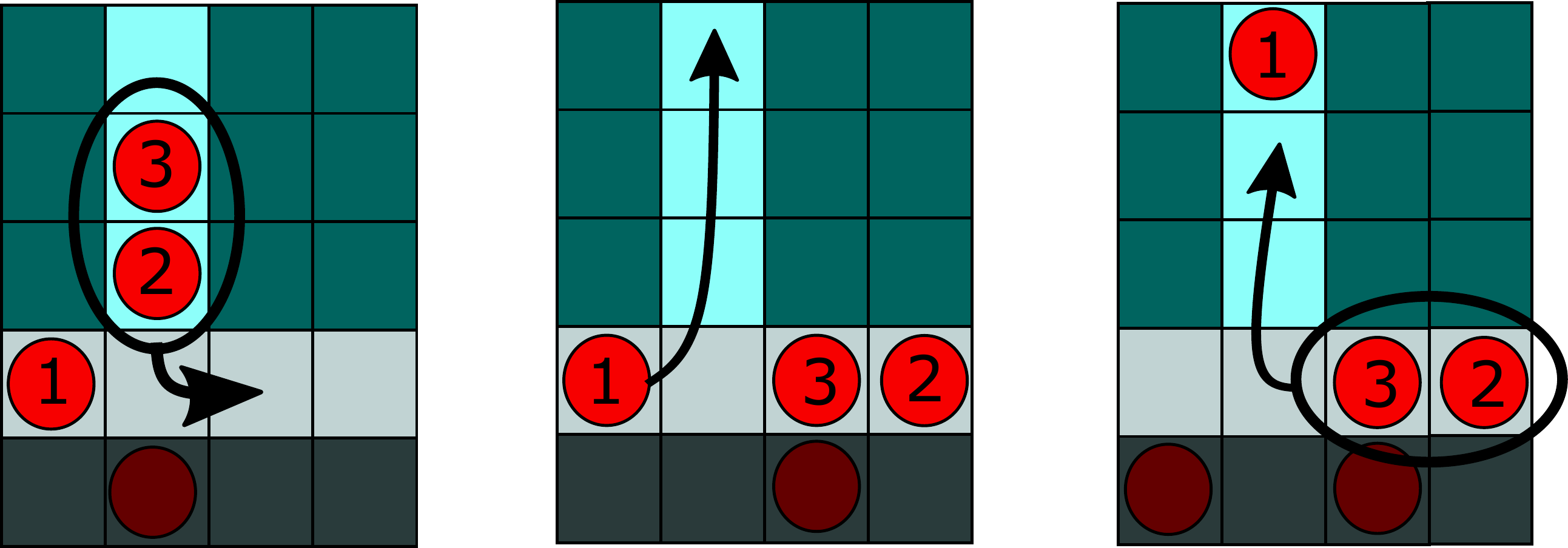}\label{fig:interf32}}\hfill
		\subfloat[Type 3]{\includegraphics[height=1in]{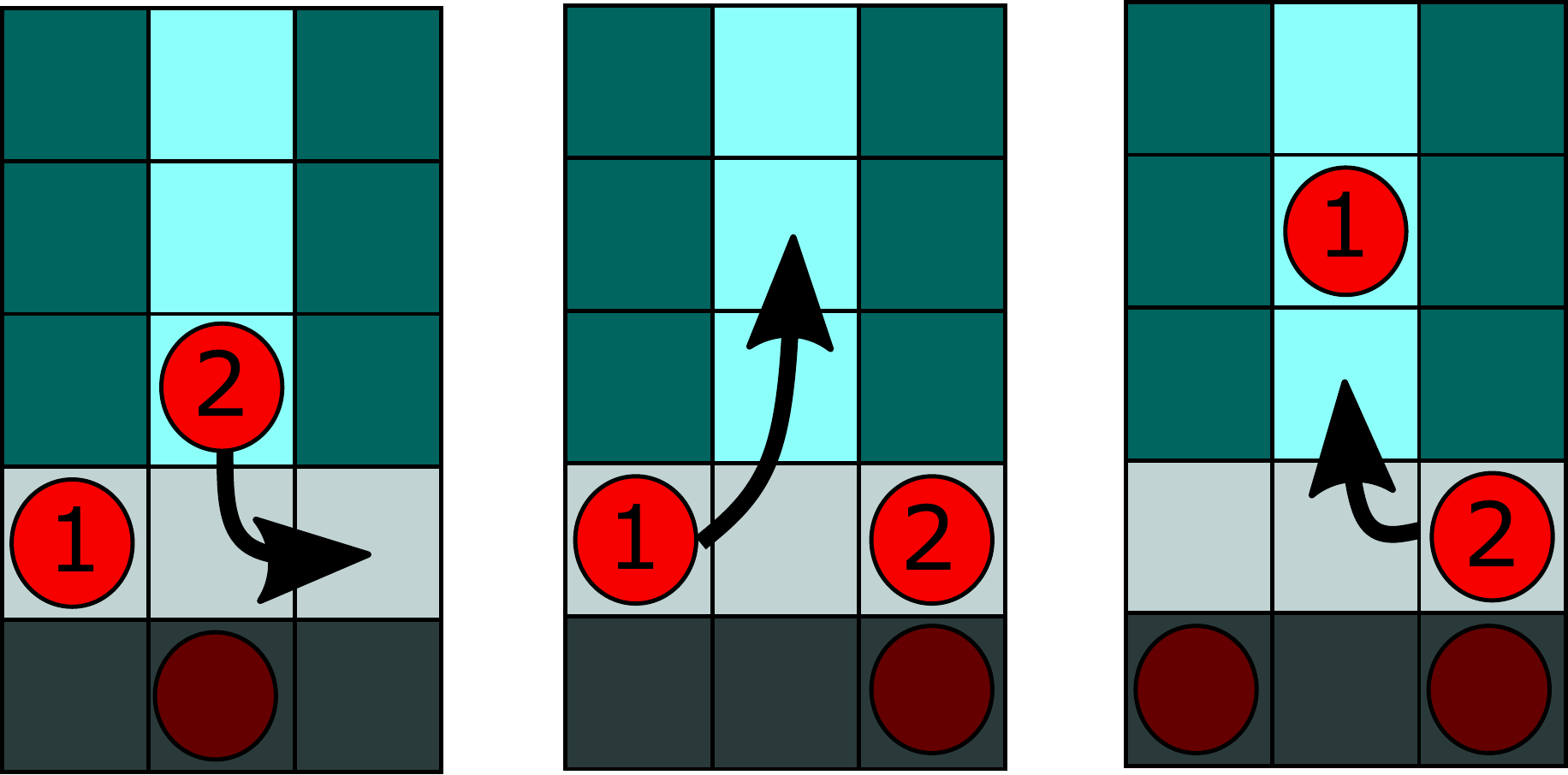}\label{fig:interf31}}\\
		\subfloat[Type 4 (PrP cabin)]{\includegraphics[height=1in]{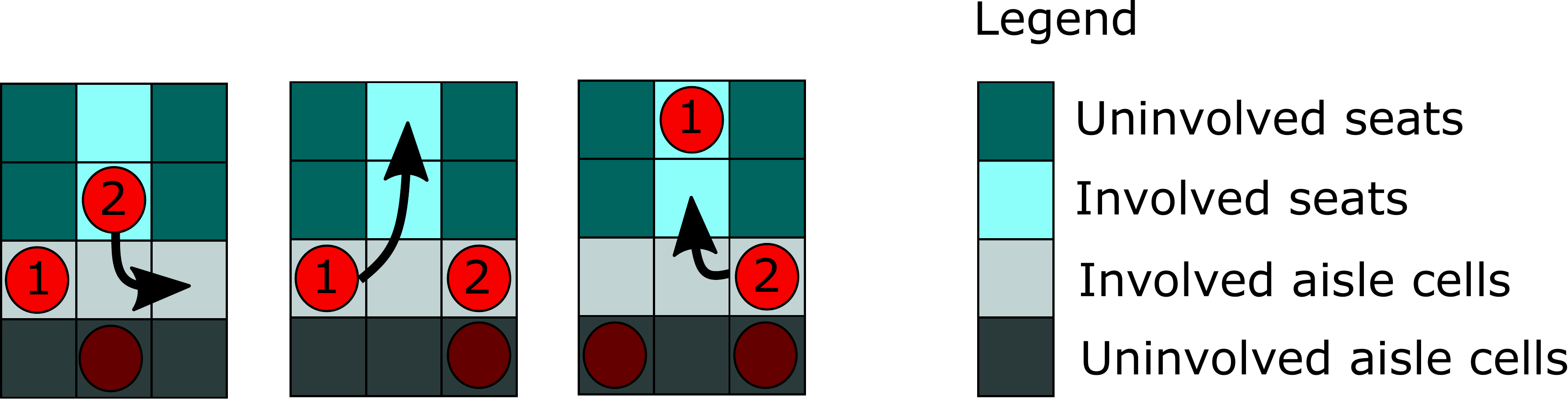}\label{fig:interf21}}%
		\caption{Some seat interference types handled by SimBaD}
		\label{fig:interf}
	\end{figure}

	Seat interference (SI) represents a major source of uncertainty during boarding phases. It happens when a passenger cannot enter freely in his row because another sat passenger hinders his path. 
	SimBaD explicitly models this phenomenon, as shown in Fig.~\ref{fig:interf}, where some cases are illustrated (for literature nomenclature (Type 1, 2, etc.), see for instance \cite{Milne2019, Schultz2018c}).
	
	To illustrate SimBaD seat interference handling, consider the Type 3 interference shown in Fig. \ref{fig:interf31}.
	The simulation of SI involves only the light-shadowed cells of Fig.\ref{fig:interf}, using the same color legend of Fig. \ref{fig:discret}. The three light-shadowed aisle cells are dedicated to the SI manoeuvre; no other passenger can use them. Passenger 1 would like to seat, after posing his luggage, and moves to the left-side cell, with respect to the seat row. Passenger 2, who makes obstruction, exits and moves to the right-hand-side cell. 
	Note that in case of aisles with multiple cells in the transversal direction, only one row of aisle is involved in the SI dynamics. The other ones admit the passengers flow not to arrest, as shown in Fig. \ref{fig:interf}.
	Therefore, Passenger 1 can seat, followed by Passenger 2. Once the two are sat, the light-shadowed aisle cells are \textit{unlocked}, and the passage of other passengers can continue as usual.
	
	Passengers velocity through the seats is set assigning a holding time of $1.8\,\si{\second}$ per cell \citep{Muller2009, Jafer2017}.
	
	The SI explicit simulation represents a remarkable feature of SimBaD. Other simulation tools, presented in literature, do not consider the dynamics of this major source of delay in boarding operations. This fact may be due to the comparative, and not quantitative, nature of these works.
	
	\subsection{Deboarding Phase}
	Deboarding phase is conceptually based on the same aforementioned features. However, little differences exist.
	
	At the beginning of the deboarding simulation, a time period necessary for the Equipment positioning is used to let some passengers to stand up, take their own luggage, and to approach the target exit door, which is still closed. This approach mimics the actual dynamics of deboarding: it is a common experience that, one landed, passengers start approaching the doors, even still closed, meanwhile air-stairs or fingers are being placed. After the Equipment positioning time, doors are open and passengers can disembark from the aircraft. The DT is defined as the time elapsing between the first and the last passenger exiting from the cabin. If more doors are involved in the deboarding phase, it is assumed they are open at the same time.
	
	When passengers stand up from their seats, it is a common experience that more than one passenger can occupy a single cell. This feature is impossible to model in this framework, because of the cabin discretisation and the exclusivity of each cell, which can accept only a passenger per time. Unfortunately, not considering this aspect would lead to unrealistic DTs, even superior to BTs. 
	Empirically, a coefficient of $0.5$ on $t_{\mathrm{lug}}$ solves the problem. The correction is applied only to narrow aisle configurations, since the wider aisles are essentially independent from the $t_{\mathrm{lug}}$, as will be clear in Section \ref{sec:results}.

	\subsection{Validation}\label{sec:validation}
\begin{figure}[hbtp]
		\centering
		\includegraphics[width=3.25in]{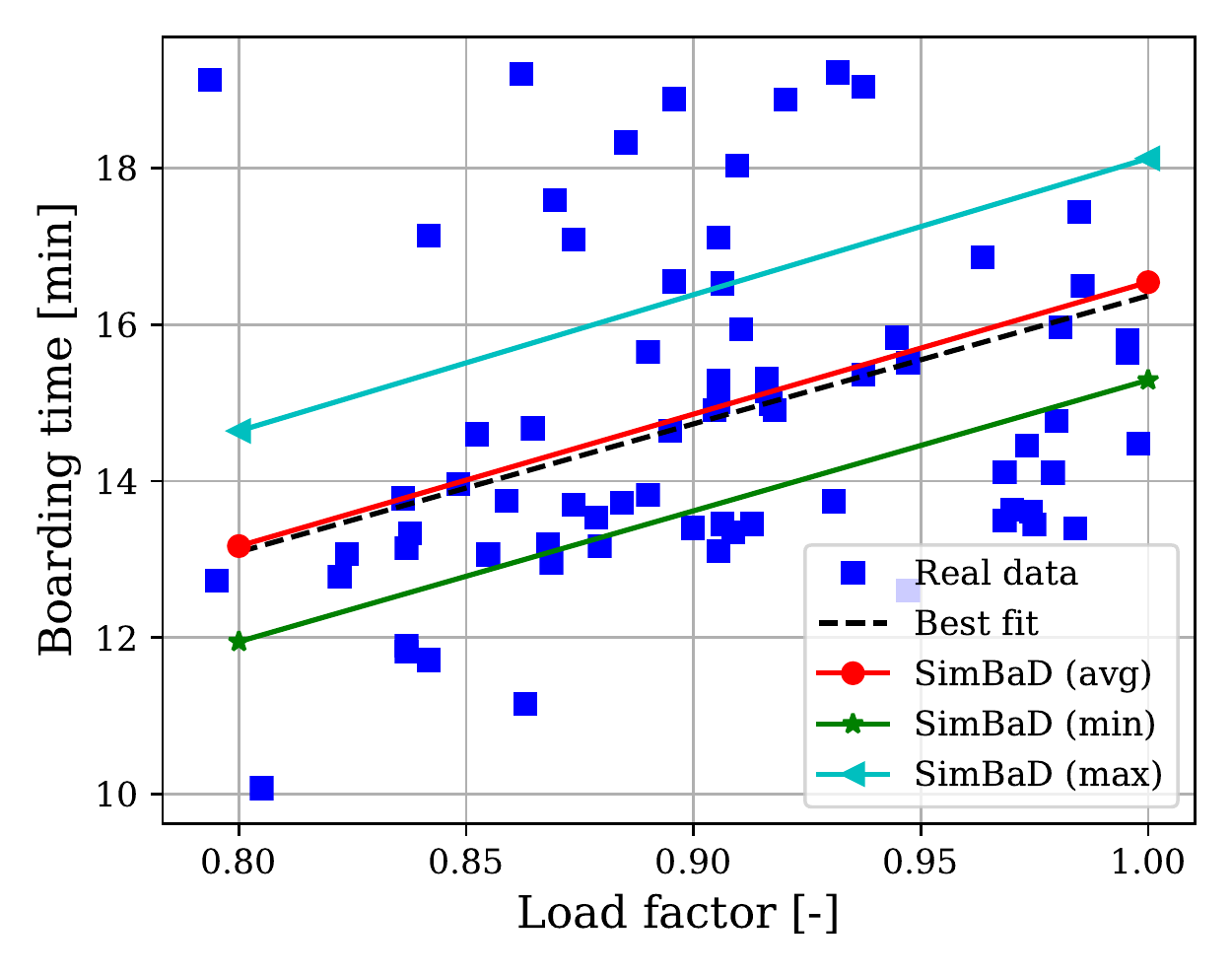}
		\caption{Validation with real data from \citep{Schultz2018} for B737/A320 aircraft ($n_{\mathrm{pax}}^{\mathrm{max}}=189$, 1 door)}
		\label{fig:valid737}
	\end{figure}

	SimBaD validation is based on the comparison between predictions and available data for different class of aircraft. In particular, B737, A320, B767 and A330 aircraft have been considered, belonging to the ICAO categories C and D.
	Figure  \ref{fig:valid737} shows real data of BTs, for B737 and A320 aircraft, taken from \citep{Schultz2018}. The data have been recorded by the authors in German airports with different airlines during the summer periods 2010–2015. The attention has been focused on the range $\mathrm{LF} \geq 0.8$.
	The same Figure shows the prediction of SimBaD for a conventional cabin layout (3+3 seats abreast), at the same conditions of \citep{Schultz2018}, i.e. $n_{\mathrm{pax}}^{\mathrm{max}}=189$ and one active door (door L1) for the boarding procedure. Of course, no overtaking is taken into account due to the conventional narrow aisle. A $t_{\mathrm{lug}}$ distribution of type B (Tab. \ref{tab:weib}) has been assumed. Each plotted point for the SimBaD simulations is the average of $200$ simulations, with random passengers-seat assignments and random passengers order of entrance. Figure \ref{fig:valid737} shows the average curve (very close to the best-fit curve of real data) and the scatter band of simulated BTs.
	It is noteworthy that for the B737-800, the manufacturer declares a BT of $15$ min for $177$ passengers boarding from the L1 door \citep{b737}, corresponding to $\mathrm{LF}=0.93$. It is possible to appreciate, also in this case, the good prediction of SimBaD simulations by looking at Fig. \ref{fig:valid737}.
	
	Other validation simulations have been performed by comparing the BTs and DTs declared by manufacturers with those retrieved by SimBaD. 
	In \citep{a330}, Airbus declares a BT of $10$ min for a A330 aircraft having $n_{\mathrm{pax}}^{\mathrm{max}}=300$, when boarding is performed through two doors. SimBaD simulations estimate the BT in $9:50\,\si{\minute}$, with a scatter range of $8:20-11:15\,\si{\minute}$, considering $t_{\mathrm{lug}}$ distribution of type A (Table \ref{tab:weib}). It is in fact reasonable that, for such class of aircraft performing medium-long journeys, every passenger carries a piece of luggage in the cabin. In the same spirit, Boeing \citep{b767} declares a BT of $13\,\si{\minute}$ for a $261$ passengers B767-200 having $n_{\mathrm{pax}}^{\mathrm{max}}=300$ when boarding is performed through L1 door. This case corresponds to $\mathrm{LF}=0.87$. SimBaD simulations estimate a BT of $12:57$ considering again $t_{\mathrm{lug}}$ distribution of type A (Table \ref{tab:weib}). The scatter range is $12:05 - 14:45$ min. For all these cases, each result is the average of $200$ simulations, with random passengers-seat assignments and random passengers order of entrance.
	
	Similar considerations can be extended to the validation of the deboarding phase. Unfortunately, a collection of real data, for a data-driven calibration, has not been found in the literature. 
	Therefore, the validation has been done only by comparing data declared by manufacturers with those recovered by SimBaD simulations, as done for the boarding validation. The same assumptions considered for the boarding phase are considered valid for the deboarding phase (e.g. luggage storing/retrieving time distributions, LFs). For this phase, passengers dislocation in the cabin is assumed from the results of boarding simulations. For deboarding simulations, each result is the average of $200$ simulations, with passengers distribution within the cabin assumed from a random boarding process (at the same conditions). It is assumed that passengers can stand up, retrieve luggage and approach exit doors from the beginning of the simulation. However, doors are open (and passengers can start leaving the cabin) after two minutes, so to simulate the time needed to install the stairs equipment.\\
	Tables \ref{tab:valid} and \ref{tab:validdeb} compare the declared and simulated BTs and DTs, respectively, for different aircraft classes, according to the loading conditions and layouts considered by the manufacturers. There are also listed the average SimBaD results for fully-loaded cabin (LF=1).
	
	\begin{table}
		\centering
		\caption{SimBaD simulations and declared BTs for different aircraft classes and scenarios}
		\label{tab:valid}
		\begin{tabular*}{\textwidth}{llllllll}
			\toprule
			{Aircraft} &  {$n_{\mathrm{pax}}^{\mathrm{max}}$} &{$t_{\mathrm{lug}}$ type} & {LF} & {N doors} & {Declared BT} & {SimBaD BT (avg)} & {SimBaD BT (avg, LF=1)}\\
			{} &  {} &  {} & {} & {} &{$[\si{\minute}]$} &{$[\si{\minute}]$}  &{$[\si{\minute}]$} \\
			\midrule
			{B737} &  {$189$} & {B} & {$0.85$} & {$1$} & {$14:00$}  & {$13:55$} & {$16:15$} \\
			{B737} &  {$189$} & {B} & {$0.94$} & {$1$} & {$15:00$}  & {$15:11$} & {$16:15$} \\
			{A320} &  {$189$} & {B} & {$0.94$} & {$2$} & {$7:30$}  & {$7:55$} & {$8:30$} \\
			{B767} &  {$300$} & {A} & {$0.87$} & {$1$} & {$13:00$} & {$12:57$} & {$15:05$}\\
			{A330} &  {$300$} & {A} & {$1$} & {$2$} & {$10:00$} & {$9:50$} &{$9:50$}\\
			\bottomrule
		\end{tabular*}
	\end{table} 
	
	\begin{table}
		\centering
		\caption{SimBaD simulations and declared DTs for different aircraft classes and scenarios}
		\label{tab:validdeb}
		\begin{tabular*}{\textwidth}{llllllll}
			\toprule
			{Aircraft} &  {Max Pax} &{$t_{\mathrm{lug}}$ type} & {LF} & {N doors} & {Declared DT} & {SimBaD DT (avg)} & {SimBaD DT (avg, LF=1)}\\
			{} &  {} &  {} & {} & {} &{$[\si{\minute}]$} &{$[\si{\minute}]$}  &{$[\si{\minute}]$} \\
			\midrule
			{B737} &  {$189$} & {B} & {$0.85$} & {$1$} & {$9:00$}  & {$9:00$} & {$10:20$} \\
			{B737} &  {$189$} & {B} & {$0.94$} & {$1$} & {$10:00$}  & {$9:50$} & {$10:20$} \\
			{A320} &  {$189$} & {B} & {$0.94$} & {$2$} & {$5:00$}  & {$5:00$} & {$5:15$} \\
			{B767} &  {$300$} & {A} & {$0.87$} & {$1$} & {$10:30$}  & {$10:40$} & {$12:20$}\\
			{A330} &  {$300$} & {A} & {$1$} & {$2$}    & {$6:00$}  & {$6:00$}  & {$6:00$}\\
			\bottomrule
		\end{tabular*}
	\end{table}

	\color{black}

	\section{Case Study: the PARSIFAL PrandtlPlane Turnaround Time}\label{sec:tat}
	
\begin{figure}[hbtp]
		\centering
		\subfloat[Comparison between conventional and PARSIFAL PrP aisles]{\includegraphics[width=0.48\columnwidth]{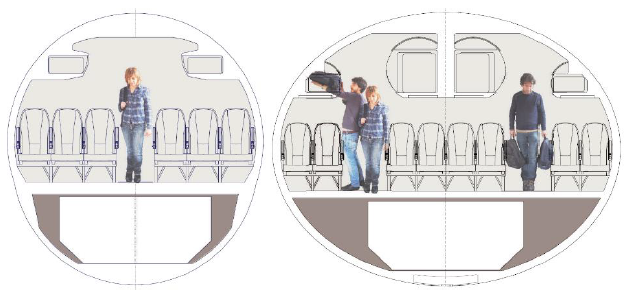}\label{fig:2pax}}%
		\hfill
		\subfloat[Comparison between conventional and PARSIFAL PrP deboarding paths]{\includegraphics[width=0.48\columnwidth]{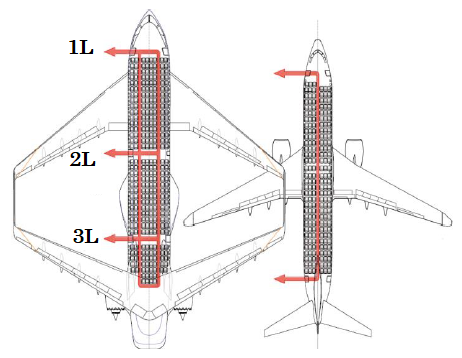}\label{fig:3doors}}\\
		\subfloat[Conceptual schemes for conventional and PARSIFAL PrP cargo bay]{\includegraphics[width=0.60\columnwidth]{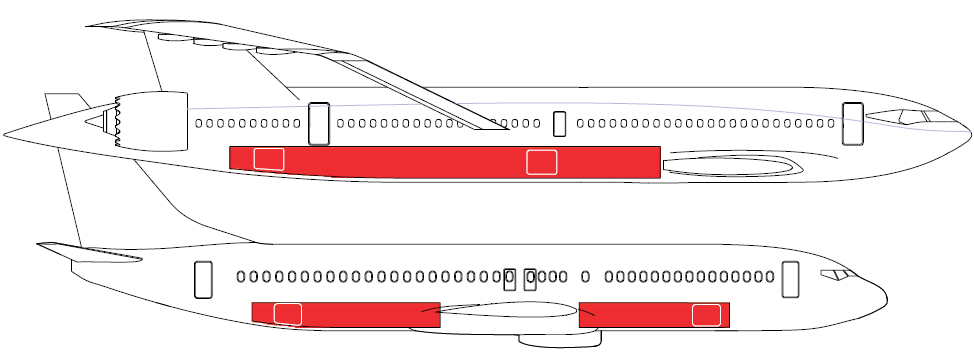}\label{fig:prpstiva}}
		\caption{PARSIFAL PrP configuration}
		\label{fig:parsifal}
	\end{figure}

	As a case study of interest, the approach is applied to the PrandtlPlane (PrP) aircraft developed in the frame of PARSIFAL Project \citep{Frediani2019}. The PrandtPlane is the application into aeronautical engineering of Prandtl's "Best Wing System" concept \citep{Prandtl1924}, i.e. the lifting system with minimum induced drag for given generated lift and wingspan. For the PARSIFAL PrP, the wingspan is limited to $36\,\si{\meter}$, in compliance to ICAO C category.
	The PARSIFAL PrP cabin architecture can host $n_{\mathrm{pax}}^{\mathrm{max}}=308$ passengers and adopts an innovative fuselage layout, with 2-4-2 passengers abreast. Two wider aisles allow two passengers to pass by, while one of them is storing/taking the luggage, as shown in Fig.~\ref{fig:2pax}. It is noteworthy that up to three doors are available for boarding/deboarding procedures (Fig. \ref{fig:3doors}). Moreover, the wing position allows for a uninterrupted cargo bay, which may host 12 LD-45 containers, as shown in Fig. \ref{fig:prpstiva}.
	Even though a horizontally-enlarged fuselage cross-section introduces many drawbacks from the structural viewpoint, as pointed out in \citep{PicchiScardaoni2019f, PicchiScardaoni2019g, PicchiScardaoni2017}, it allows exploiting the enhanced payload capacity of the PrP, deriving from a superior aerodynamic efficiency if compared to competitors.
	
	\subsection{SimBaD Simulations for PrP}\label{sec:results}
	\subsubsection{Effect of Wider Aisle}
	The net effect of a wider aisle, with respect to conventional aircraft, has been primarily investigated. For this purpose, both the PARSIFAL PrP cabin layout (Fig.  \ref{fig:prpdiscrete}) and an alternative cabin with the same layout of the PARSIFAL PrP one but with conventional narrow aisles have been considered. Hereafter the PrP with wide aisles is referenced as \textit{wide-PrP}, whilst the PrP with narrow conventional aisles is referenced as \textit{narrow-PrP}.
	
\begin{figure}[hbtp]
		\centering
		\includegraphics[width=3.25in]{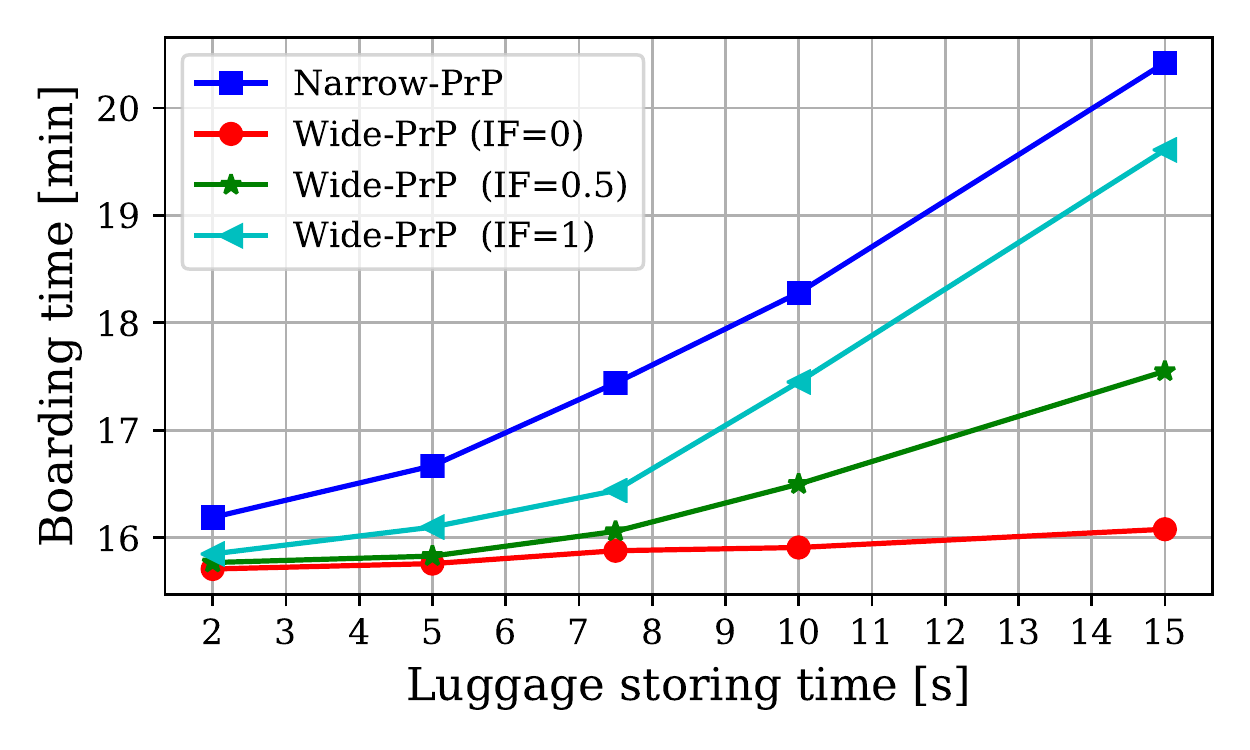}
		\caption{SimBaD simulations}
		\label{fig:PRP1DN}
	\end{figure}

	Considering the passengers velocity distribution of Tab. \ref{tab:weib}, all other parameters being constant, a sensitivity study  has been performed with respect to the luggage storing time $t_{\mathrm{lug}}$, constant for all passengers, and to the $\mathrm{IF}$. Results are shown in Fig.  \ref{fig:PRP1DN}. In this analysis only one active door, namely 1L door, is active, in order to eliminate the effect of multiple doors. For a \textit{perfect} wide-PrP, i.e. when $\mathrm{IF}=0$, the BT is remarkably independent from $t_{\mathrm{lug}}$. This is due to the ability of the wide-PrP to not to stop the passengers flow.  
	On the contrary, the narrow-PrP configuration suffers much more as the $t_{\mathrm{lug}}$ becomes important. Of course, the wide-PrP avoids bottlenecks and transfers serial activities into parallel activities, as much as possible. Conversely, it does not happen in the narrow-PrP configuration, where bottlenecks always stop the passengers flow.
	More likely, intermediate IFs, due to the wide range of human behaviours, are more representative of the reality. The wide-PrP configuration seems to be substantially independent from $t_{\mathrm{lug}}$ for IFs less than $0.5$. 
	In Fig.  \ref{fig:PRP1DN}, the wide-PrP case for $\mathrm{IF}=1$ is not equal to the narrow-PrP case because in the former case the $\mathrm{IF}$ acts only on the storing luggage events, whilst during seat interference cases the passengers flow is allowed to continue. Of course, with narrow aisles, it cannot happen, by default.
	
	The fact that the ideal PrP is essentially independent from  $t_{\mathrm{lug}}$ means that PrP BT depends only on the speed of passengers movement inside the cabin. Airliners can exploit this observation in a convenient manner and adapt their strategies accordingly. In fact, if the BT becomes independent from the adopted luggage politics, airliners can rely on a much more flexible aircraft when it comes to on-board luggage. This translates into commercial opportunities resulting from such a flexibility. For instance, airliners could assess the possibility of allowing passengers to always store one (or even two) trolley in the overhead bins, and they could think to assign a specific space for the on-board luggage associated with every seat. This strategy would avoid passengers going around looking for a free space in the overhead bins when the space over their seats is occupied. The resulting benefits in terms of time efficiency are clear. 
	
	\subsubsection{Boarding and Deboarding Simulations}
\begin{figure}[hbtp]
		\centering
		\subfloat[1 door]{\includegraphics[width=0.48\columnwidth]{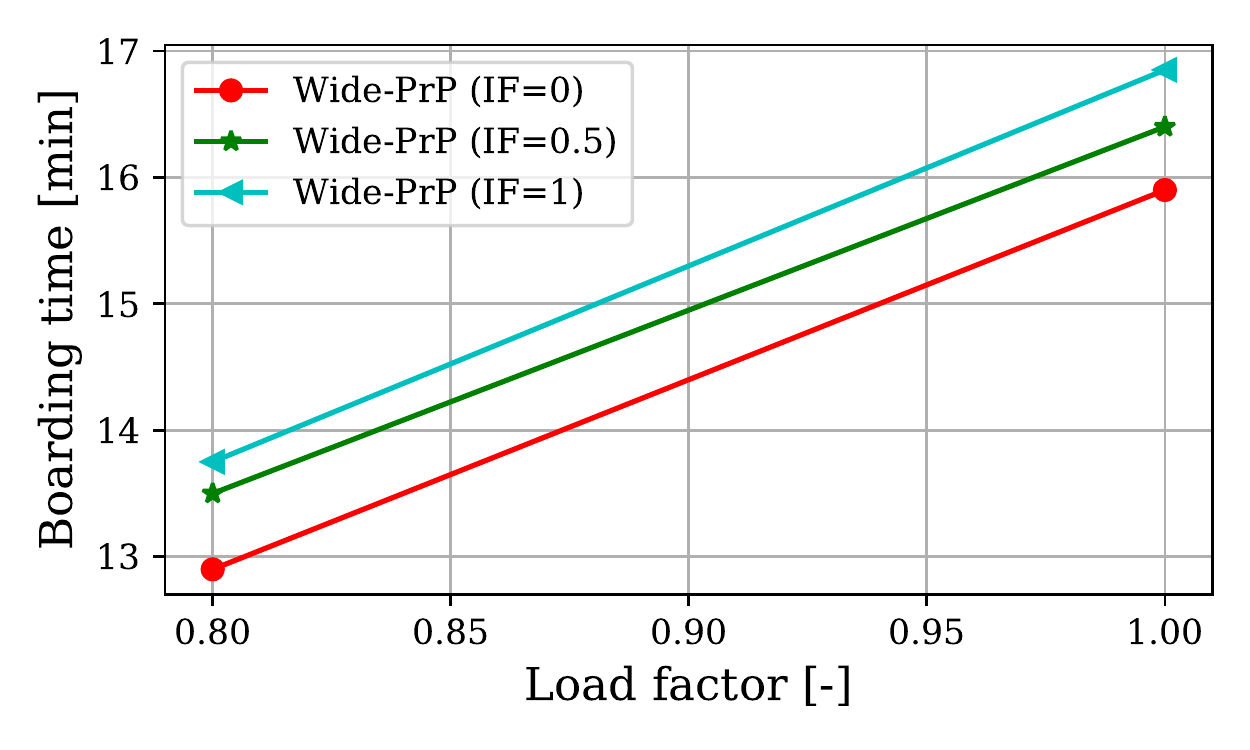}\label{fig:PRP1D}}%
		\subfloat[2 doors]{\includegraphics[width=0.48\columnwidth]{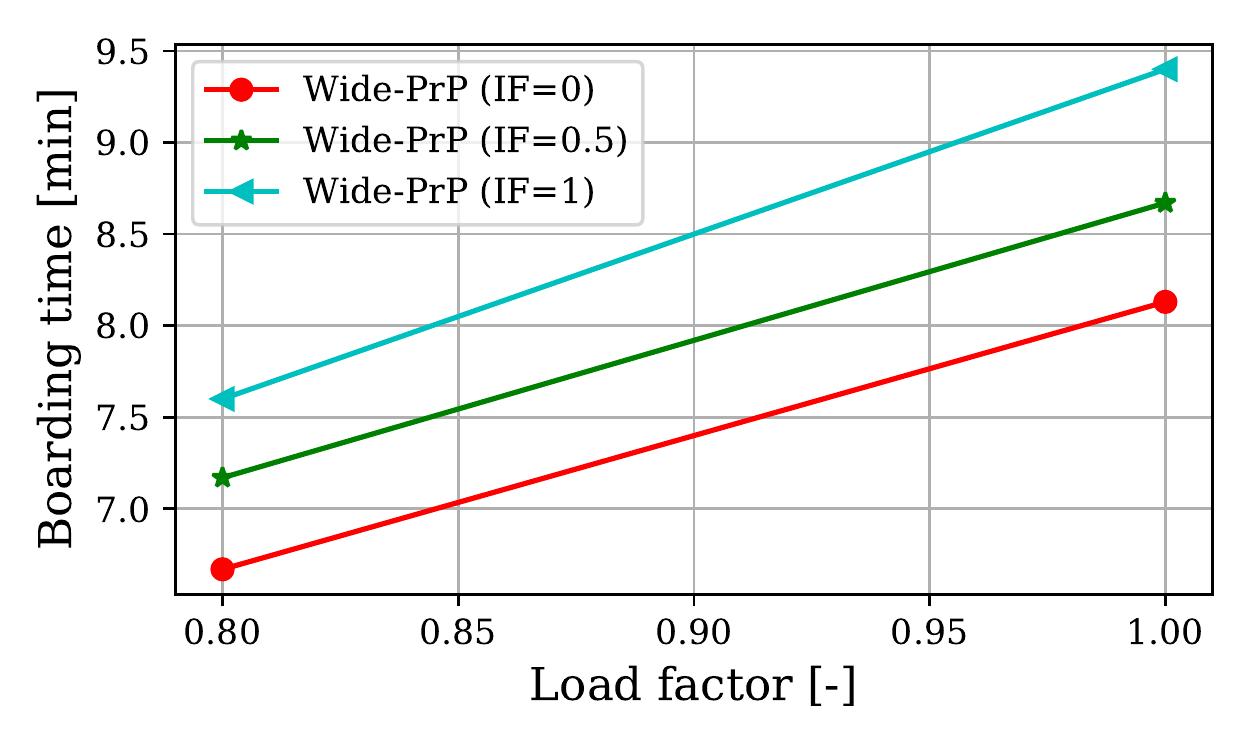}\label{fig:PRP2D}}%
		\hfill
		\subfloat[3 doors]{\includegraphics[width=0.48\columnwidth]{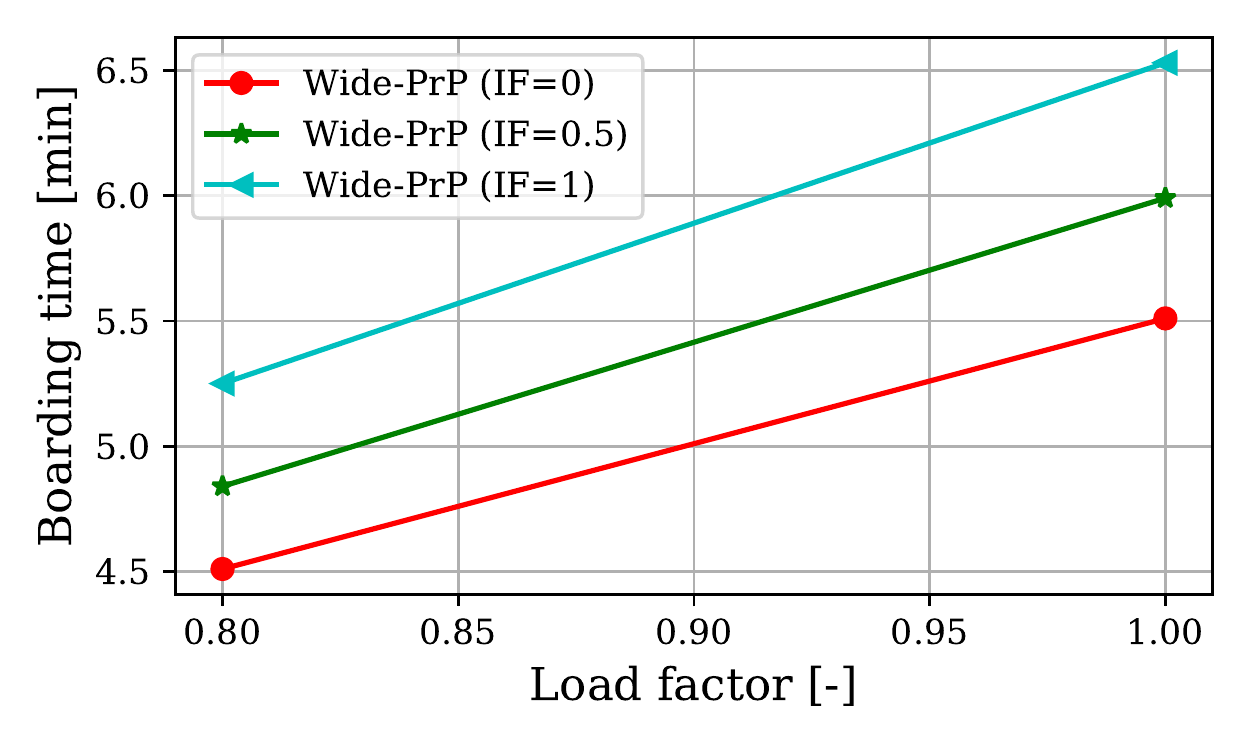}\label{fig:PRP3D}}
		\caption{PARSIFAL PrP average boarding time (SimBaD simulations)}
		\label{fig:btparsifal}
	\end{figure}

\begin{figure}[hbtp]
		\centering
		\includegraphics[width=3.25in]{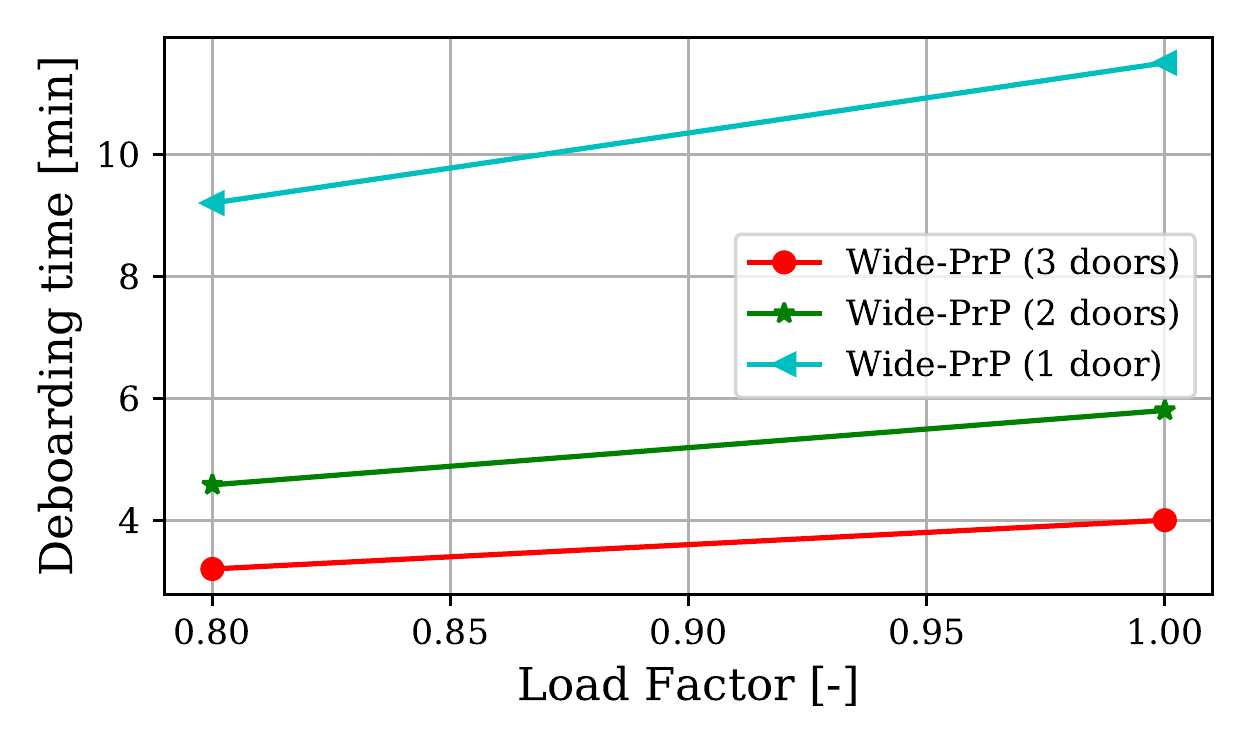}
		\caption{PARSIFAL PrP average deboarding time (SimBaD simulations)}
		\label{fig:prpdt}
	\end{figure}

	Figs. \ref{fig:PRP1D},  \ref{fig:PRP2D} and \ref{fig:PRP3D} show the sensitivity studies of the BT with respect to the IF and to the LF of the PARSIFAL PrP configuration, considering one, two or three active doors, respectively. Similarly, Fig. \ref{fig:prpdt} shows the simulation of the DT.
	A luggage politic of type A (see Tab. \ref{tab:weib}) is henceforth assumed. The use of one door can be interpreted as the use of the finger.
	
	The number of active doors is very effective on the resulting BTs and DTs. In particular, the (de)boarding via two active doors is by far more efficient than the single door counterpart. It is noteworthy that the activation of a third door reduces the absolute time saving with respect to the activation of a second door, thus suggesting a saturation effect as the number of doors increases.
	It is observed that the IF affects the resulting BTs for up to $1.5\,\si{\min}$.
	
	\begin{table}
		\centering
		\caption{PARSIFAL PrP boarding and deboarding rates}
		\label{tab:rates}
		\begin{tabular*}{\textwidth}{lllllll}
			\toprule
			{Aircraft} &  {Doors} & {$n_{\mathrm{pax}}^{\mathrm{max}}$} &{LF}& {IF} &{Boarding rate} & {deboarding rate}\\
			& &  && &[pax/min/door]& [pax/min/door]\\
			\midrule
			{PARSIFAL PrP} &  1&308&1&0.5&18.7&26.8\\
			{PARSIFAL PrP} &  2&308&1&0.5&17.7&26.5\\
			{PARSIFAL PrP} &  3&308&1&0.5&17.1&25.7\\
			\bottomrule
		\end{tabular*}
	\end{table} 
	Table \ref{tab:rates} shows the boarding and deboarding rates obtained from SimBaD simulations. These values are used in the following Section to estimate the PARSIFAL PrP TAT.

	\subsection{PARSIFAL PrP Turnaround Time Estimation}
	To estimate the PARSIFAL PrP TAT, two operating scenarios are considered, which are representative of actual operating conditions. According to manuals (for instance, \cite{a320}), they are the \textit{Full service} and the \textit{Outstation} scenario, which provide the theoretical maximum and minimum TAT for an aircraft.
	
	The Full-service scenario is characterised as follows:
	\begin{enumerate}
		\item passengers move in and out the airplane through one single door (use of the finger);
		\item maximum supply of fuel (passengers are not allowed on board during refuelling according to EU-OPS 1.305 (FAR 121.570)  \citep{EC2008};
		\item maximum supply of provisions (catering).
	\end{enumerate}
	Conversely, the Outstation scenario is characterised as follows:
	\begin{enumerate}
		\item passengers embark and disembark the airplane through 2 doors, assuming an equal repartition;
		\item no refuelling;
		\item minimum supply of provisions (catering).
	\end{enumerate}
	In both scenarios the cargo exchange, i.e. the number of containers unloaded and loaded, is the maximum possible.

	Both scenarios are considered separately in the estimation of the PARSIFAL PrP TAT, and are compared to the counterpart for the A320 cases. The PARSIFAL PrP is in fact designed to be compliant with ICAO C class.
	Considering the greater number of PARSIFAL PrP transported passengers, catering and potable water supply have been increased proportionally to the ratio $308/189$, $308$ being the maximum number of passengers of PARSIFAL PrP and $189$ being the counterpart for the A320 aircraft.
	The number of $12$ containers and the quantity of fuel, $27$ tons, derive from preliminary design of the aircraft \citep{Cipolla2018a, Cipolla2020}.
	
	Table \ref{tab:datatat} shows the values used for the TAT evaluations. The TAT of the PARSIFAL PrP and the A320 are estimated and compared in  the Full Service and Outstation operating conditions. All of the listed values and assumptions, from \citep{a320}, when possible, are considered valid also for the PARSIFAL PrP case. 
	
	\begin{table}
		\small
		\centering
		\caption{Definition of the reference conditions for the calculation of turnaround time }
		\label{tab:datatat}
		\begin{tabular*}{\textwidth}{lllll}
			\toprule
			& \multicolumn{2}{c}{Full Service	
			}& \multicolumn{2}{c}{Outstation}\\
			\midrule
			& A320 & PARSIFAL PrP & A320 & PARSIFAL PrP\\
			\midrule
			PASSENGER HANDLING &&&&\\
			\midrule
			Passengers$^1$  &$150$ (LF=0.8)&$244$ (LF=0.8)&$180$ (LF=0.95)&$293$ (LF=0.95)\\
			Doors  &1&1&2&2\\
			Passengers per door  & 100\% & 100\% & 50\% & 50\% \\
			Equipment positioning $[\si{\min}]$ & 2&2&2&2\\
			Equipment removal  $[\si{\min}]$ &   2&2&2&2\\
			Boarding rate [pax/min/door]  &12&	18.7&	12&	17.7\\
			LSR + headcounting  $[\si{\min}]$  & 2&2&2&2\\
			Deplaning rate [pax/min/door]  &20&	26.8&	18&	26.5
			\\
			\midrule
			CARGO &&&&\\
			\midrule
			Equipment positioning  $[\si{\min}]$  & 2&2&2&2\\
			Equipment removal  $[\si{\min}]$ & 1.5&1.5&1.5&1.5\\
			100\% cargo exchange (baggage only)  &\checkmark &\checkmark &\checkmark &\checkmark\\
			FWD cargo compartment containers   &3	&6	&3	&6
			\\
			AFT cargo compartment containers  &4	&6	&4&	6
			\\
			Container unloading rate [min/container]  & 1.5&1.5&1.5&1.5\\
			Container loading rate [min/container]  & 1.5&1.5&1.5&1.5\\
			\midrule
			REFUELING &&&&\\
			\midrule
			Fuel quantity [m$^3$]  &20&33.7&-&-\\
			Fuel flow [m$^3$/min]  &1.25&1.25&-&-\\
			Truck positioning/removal [min]  &2.5&2.5&-&-\\
			Truck connection/disconnection [min]  &2.5&2.5&-&-\\
			Passengers admitted on board  &NO&NO&-&-\\
			\midrule
			CATERING &&&&\\
			\midrule
			Catering trucks   &1&1&1&1\\
			Equipment positioning [min]   &2&2&2&2\\
			Equipment removal [min]  &1.5&1.5&1.5&1.5\\
			Time to drive from doors [min]  &2&2&2&2\\
			Full Size Trolley Equivalent (FSTE) door 1R  &4&7&1&2\\
			FSTE: door 4R  &7&11&0&0\\
			Time for trolley exchange [min/FTSE]  &1.2&1.2&1.2&1.2\\
			Minimum time for catering [min]  &-&-&3.5&3.5\\
			\midrule
			CLEANING &\multicolumn{4}{c}{All available time}\\
			\bottomrule
		\end{tabular*}
		$^1$ For the A320, it is assumed $n_{pax}^{max}=189$, whilst the number of passengers is taken from \cite{a320}. For the PARSIFAL PrP case, $n_{pax}^{max}=308$, and the same LF of the A320 counterpart is assumed, thus leading to the actual number of passengers involved in the scenarios.
	\end{table}
	
	\begin{figure*}[hbtp]
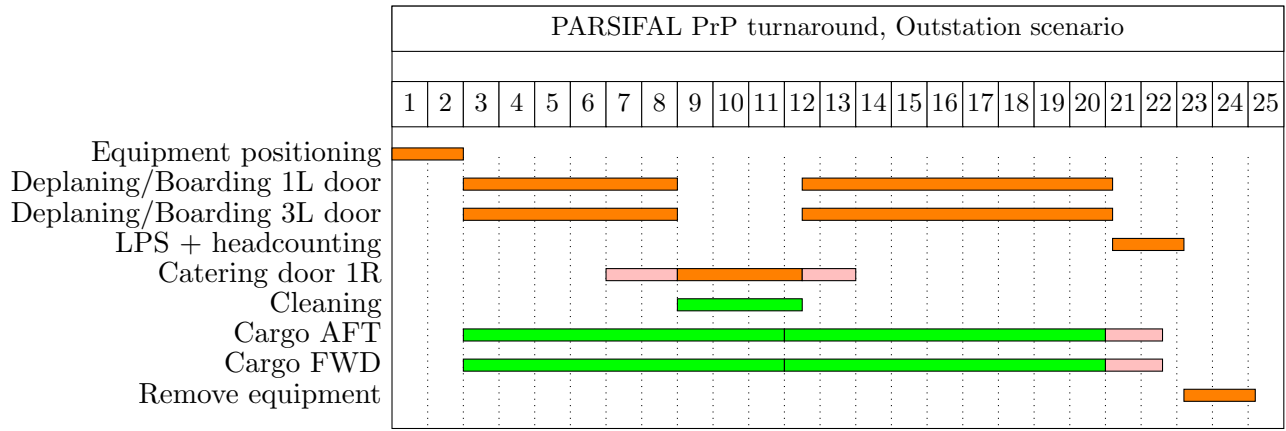

		\centering
		\begin{ganttchart}[x unit=0.01cm,
			expand chart=\textwidth, vgrid={*9{draw=none}, dotted}, bar top shift=-0.3, bar height=0.4, y unit chart=0.4cm]{1}{250}
			\gantttitle{PARSIFAL PrP turnaround, Outstation scenario}{250} \\
			\gantttitlelist{1,...,25}{10} \\
			\ganttbar[bar/.append style={fill=orange}]{Equipment positioning}{1}{20} \\
			\ganttbar[bar/.append style={fill=orange}]{Deplaning/Boarding 1L door}{21}{80} 
			\ganttbar[bar/.append style={fill=orange}]{}{116}{202} \\
			\ganttbar[bar/.append style={fill=orange}]{Deplaning/Boarding 3L door}{21}{80} 
			\ganttbar[bar/.append style={fill=orange}]{}{116}{202} \\
			\ganttbar[bar/.append style={fill=orange}]{LPS + headcounting}{203}{222} \\
			\ganttbar[bar/.append style={fill=pink}]{Catering door 1R}{61}{80}\ganttbar[bar/.append style={fill=orange}]{}{81}{115}\ganttbar[bar/.append style={fill=pink}]{}{116}{130}\\
			\ganttbar[bar/.append style={fill=green}]{Cleaning}{81}{115}\\
			\ganttbar[bar/.append style={fill=green}]{Cargo AFT}{21}{110}\ganttbar[bar/.append style={fill=green}]{}{111}{200}\ganttbar[bar/.append style={fill=pink}]{}{201}{216}\\
			\ganttbar[bar/.append style={fill=green}]{Cargo FWD}{21}{110}\ganttbar[bar/.append style={fill=green}]{}{111}{200}\ganttbar[bar/.append style={fill=pink}]{}{201}{216}\\
			\ganttbar[bar/.append style={fill=orange}]{Remove equipment}{223}{242}
		\end{ganttchart}
		\caption{Turnaround process for PARSIFAL PrP, Outstation scenario. In orange the critical path, in pink the positioning activities, in green the non-critical activities. Time in minutes.}
		\label{fig:tatprpout}
	\end{figure*}

	\begin{figure*}[hbtp]
		\centering
		\begin{ganttchart}[x unit=0.01cm,
			expand chart=\textwidth, vgrid={*9{draw=none}, dotted}, bar top shift=-0.3, bar height=0.4, y unit chart=0.4cm]{1}{560}
			\gantttitle{PARSIFAL PrP turnaround, Full-service scenario}{560} \\
			\gantttitlelist{2,4,6,8,10,12,14,16,18,20,22,24,26,28,30,32,34,36,38,40,42,44,46,48,50,52,54,56}{20} \\
			\ganttbar[bar/.append style={fill=orange}]{Equipment positioning}{1}{20} \\
			\ganttbar[bar/.append style={fill=orange}]{Deplaning/Boarding 1L door}{21}{110} 
			\ganttbar[bar/.append style={fill=orange}]{}{383}{514} \\
			\ganttbar[bar/.append style={fill=orange}]{LPS + headcounting}{515}{534} \\
			\ganttbar[bar/.append style={fill=pink}]{Catering door 1R}{91}{110}\ganttbar[bar/.append style={fill=orange}]{}{111}{195}\ganttbar[bar/.append style={fill=pink}]{}{196}{215}\\
			\ganttbar[bar/.append style={fill=pink}]{Catering door 3R}{216}{235}
			\ganttbar[bar/.append style={fill=pink}]{}{236}{250}
			\ganttbar[bar/.append style={fill=orange}]{}{251}{382}\ganttbar[bar/.append style={fill=pink}]{}{383}{397}
			\\
			\ganttbar[bar/.append style={fill=green}]{Cleaning}{111}{382}\\
			\ganttbar[bar/.append style={fill=green}]{Cargo AFT}{21}{110}\ganttbar[bar/.append style={fill=green}]{}{111}{200}\ganttbar[bar/.append style={fill=pink}]{}{201}{216}\\
			\ganttbar[bar/.append style={fill=green}]{Cargo FWD}{21}{110}\ganttbar[bar/.append style={fill=green}]{}{111}{200}\ganttbar[bar/.append style={fill=pink}]{}{201}{216}\\
			\ganttbar[bar/.append style={fill=pink}]{Refuelling}{86}{110}
			\ganttbar[bar/.append style={fill=green}]{}{111}{380}
			\ganttbar[bar/.append style={fill=pink}]{}{381}{405}\\
			\ganttbar[bar/.append style={fill=orange}]{Remove equipment}{535}{554}
		\end{ganttchart}
		\caption{Turnaround process for PARSIFAL PrP, Full-service scenario. In orange the critical path, in pink the positioning activities, in green the non-critical activities. Time in minutes.}
		\label{fig:tatprpfull}
	\end{figure*}
	
	Figures \ref{fig:tatprpout} and \ref{fig:tatprpfull} show the Gantt diagrams of the PARSIFAL PrP turnaround process. Table \ref{tab:tats} summarises the results of the comparison between the PARSIFAL PrP case and the A320. Such results provide an assessment of the trade-off between the two opposite goals, i.e., improving the payload capabilities of an aircraft while keeping the TAT as short as possible. 
	The comparison of the PARSIFAL PrP case with the A320 shows that a significantly higher number of passengers and containers (more than +50\%) is "paid" with a limited increase in TAT (up to 25\%). 
	
	\begin{table}
		\centering
		\caption{Absolute values and percent increments of passengers, containers and TAT for A320 and PARSIFAL PrP}
		\label{tab:tats}
		\begin{tabular*}{\textwidth}{lllllll}
			\toprule
			&\multicolumn{3}{c}{Full service} & \multicolumn{3}{c}{Outstation}\\
			\midrule
			& A320 & PARSIFAL PrP & Increment & A320 & PARSIFAL PrP & Increment \\
			\midrule
			Passengers & 150 & 244& +63\% &180 &293 &+63\%\\
			N Containers & 7& 12&+71\% & 7& 12&+71\%\\
			TAT [min] &44 & 55& +25\%&22 &24.5 &+11\%\\
			\bottomrule
		\end{tabular*}
	\end{table}

	\section{Discussion and Conclusions}\label{sec:conclusions}
	This paper aimed at estimating the TAT for unconventional aircraft in the preliminary design phases. In this framework, SimBaD was presented as a tool for the estimation of the boarding and deboarding times for unconventional cabin layouts. In fact, boarding and deboarding represent critical activities in the TAT process.\\
	The importance of TAT relies on the fact that it is a key performance indicator of value creation potential of a new aircraft concept, in the perspective of both the airline and the airport. As a consequence, estimating the TAT of an aircraft already in the preliminary design phases can orient the designers to make more effective choices that, in the end, may result in improved expected economic performance of the aircraft under investigation and, thereby, improved marketability potential. Predicting how the TAT is affected by the design choices can be used as a driver to move the design towards more efficient solutions. However, due to the high levels of uncertainty of the preliminary design phases, estimating the TAT is not a trivial task. Existing works focusing on this topic are a few and consider conventional single-aisle architectures. 
	
	The most challenging task is the estimation of the boarding and deboarding times. There were a few solutions in the literature to predict such times due to fuselage and cabin layout variations that may be typical of the early design phases of any aeronautical project. In this paper, the development of a simulation tool, SimBaD, allowed to simulate boarding and deboarding phases taking different delay and uncertainty sources into account. 
	
	{ As far as the model limitations are concerned, it does not take into account the saturation of overhead bins, which may deeply affect the final boarding/deboarding time \citep{Milne2014, Milne2018, Milne2016}, the effect of groups \citep{Qiang2016a, Tang2018} and passengers' individual properties \citep{Qiang2014} in boarding/deboarding, and does not implement a specific detailed model to accurately estimate ground operations.}	
	Despite its limitations, which also represent an opportunity for future studies, SimBaD seems to be a versatile tool, capable of handling different cabin layouts, thus allowing also for sensitivity and comparative analyses of the internal cabin design. { In so doing, SimBaD tries to overcome literature gaps, already discussed in Sec. \ref{sec:LR}. More in detail, if compared to other agent-based tools proposed in the literature \citep{Ferrari2005, Schmidt2016, Schultz2017a}, SimBaD is capable to handle different cabin geometries, with many aisles and doors, even unconventional ones. Moreover, it is able to simulate both boarding and deboarding, including explicit simulations of many kind of seat interferences, and including the possibility of passengers overtaking.\\	}
	SimBaD validation demonstrated the accuracy of the tool if compared to the actual times for conventional aircraft of different classes. The simulations applied to the disruptive PARSIFAL PrandtlPlane shown that such aircraft allows transporting more passengers ($+ 60\%$) and cargo if compared to its conventional competitors for the "price" of a $+25\%$ increase of its TAT. This is due to the particular design choices adopted in the preliminary design phases driven by TAT considerations. In the future, we aim at extending the potentiality of SimBaD by including the possibility to handle more sophisticated cabin layout, to insert counter-moving passengers and to simulate emergency evacuations and priority boarding strategies, besides other limitations highlighted above that, if addressed and included in the model, could boost its usefulness.  More in general, the TAT estimations can be further improved by exploiting more detailed models to estimate ground operations. For example, unconventional aircraft architecture strategies for efficient cargo unloading and loading are still to be investigated. { Also, future studies might explore the interdependence between aircraft design decisions and aircraft performance on ground integrating quantitative approaches with qualitative ones. 
		In addition to cabin layout factors, other factors such as group passengers, luggage with passenger, and luggage compartment capacity can be taken into consideration in the BT  simulation in the future. Besides, estimating the TAT with other activities during aircraft turnaround may result in more accurate previsions.}\\	
	Finally, future studies might show the importance of the turnaround time as key value creation driver in the frame of new aircraft concepts’ economic assessments.

	\section*{Aknowledgement}
	This paper presents part of the activities
	carried out within the research project PARSIFAL
	("PrandtlPlane ARchitecture for the Sustainable Improvement of Future AirpLanes"), which has been
	funded by the European Union under the Horizon 2020
	Research and Innovation Program (Grant Agreement
	n.723149).

	\FloatBarrier
	\bibliographystyle{cas-model2-names}
	\bibliography{library}   

\end{document}